\documentclass[article]{revtex4}
\usepackage{amsfonts,amsmath,amsthm}

\usepackage{amsmath,amssymb}  

\newcommand*{\AxChoice}{Axiom of Choice}
\newcommand*{\norm}[1]{\bigl|#1\bigr|}
\newcommand*{\trace}{\mathop{\mathrm{Tr}}}

\newcommand*{\scrH}{\mathord{\mathcal{H}}}%
\newcommand*{\scrC}{\mathord{\mathcal{C}}}%
\newcommand*{\scrB}{\mathord{\mathcal{B}}}%
\newcommand*{\AAA}{{\mathbb{A}}}
\newcommand*{\NN}{{\mathbb{N}}}
\newcommand*{\ZZ}{{\mathbb{Z}}}
\newcommand*{\QQ}{{\mathbb{Q}}}
\newcommand*{\RR}{{\mathbb{R}}}
\newcommand*{\CC}{{\mathbb{C}}}
\newcommand*{\scrE}{\mathord{\mathcal{E}}}%
\newcommand*{\scrT}{\mathord{\mathcal{T}}}%
\newcommand*{\scrK}{\mathord{\mathcal{K}}}%
\newcommand*{\scrN}{\mathord{\mathcal{N}}}%
\newcommand*{\scrD}{\mathord{\mathcal{D}}}%
\newcommand*{\scrP}{\mathord{\mathcal{P}}}
\newcommand*{\scrS}{\mathord{\mathcal{S}}}
\newcommand*{\Shv}{\mathop{\mathrm{S\lowercase{hv}}}}

\DeclareRobustCommand\openone{\leavevmode\hbox{\small1\normalsize\kern-.33em1}}%

\newcommand*{\EsubS}{\ensuremath{\mathord{\mathcal{E}_{\scrS}}}}
\newcommand*{\RsubC}{\ensuremath{\mathord{\RR_{\mathrm{C}}}}}
\newcommand*{\RsubD}{\ensuremath{\mathord{\RR_{\mathrm{D}}}}}

\newtheorem{theorem}{Theorem}
\newtheorem{definition}{Definition}
\newtheorem{lemma}[definition]{Lemma}
\newtheorem{proposition}[definition]{Proposition}

\newtheorem{corollary}[definition]{Corollary}

\begin{document}
\bibliographystyle{apsrev}

\title{ The Mathematical Structure of Quantum Real Numbers}

\author{John V Corbett}
\email{jvc@ics.mq.edu.au}
\affiliation{Department of Mathematics, Macquarie University, N.S.W. 2109, Australia} 

\date{\today}

\begin{abstract}
The mathematical structure of the sheaf of Dedekind real numbers $\RsubD(X)$ for a quantum system  is discussed.  The algebra of physical qualities is represented by an $O^{*}$ algebra $\mathcal M$ that acts on a Hilbert space that carries an irreducible representation of the symmetry group of the system. $X =\EsubS(\mathcal M)$, the state space for $\mathcal M$, has the weak topology generated by the functions $ a_{Q}(\cdot)$, defined for $\hat A \in  \mathcal M_{sa} $ and $\forall  \hat \rho  \in \EsubS(\mathcal M) $, by $ a_{Q}( \hat \rho) = Tr \hat A \hat \rho $.  For any open subset $W$ of  $\EsubS(\mathcal M)$, the function $ a_{Q}|_{W}$ is the numerical value of the quality $\hat A$ defined to the extent $W$. The example of the quantum real numbers for a single Galilean relativistic particle is given. 
\end{abstract}



\maketitle

\subsection{Introduction.}
The quantum real number interpretation of quantum mechanics is based on the assumption that the  numerical values taken by the attributes of a quantum system are not standard set theoretical real numbers but are topos theoretical real numbers called quantum real numbers (qr-numbers). Standard real numbers are still obtained in measurements but the "direct connection between observation properties and properties possessed by the independently existing object"\cite{cush} is cut, an indirect connection is made through the experimental measurement processes. Experimental measurement processes, as exemplified by $\epsilon$ sharp collimation\cite{durt2}, produce an approximate standard real number from an exact qr-number. The interpretation acknowledges that any experimental measurement only yields a standard real number with limited accuracy \cite{kirk} however in it each approximate standard real number corresponds to an exact qr-number.

The mathematical elements of the qr-numbers interpretation exist in standard quantum theory, it starts from von Neumann's assumption\cite{vonneumann} that every quantum system is associated with a Hilbert space $\mathcal{H}$. Then, as in the standard interpretation, the physical attributes (qualities) of the quantum system are represented by self-adjoint operators that act on vectors in $\mathcal{H}$. The operators form a $\ast$-algebra $\mathcal M$ of unbounded operators and the state space $\EsubS(\mathcal{M}) $ of the system is the space of normalized linear functionals  on $\mathcal M$. $\EsubS(\mathcal{M}) \subset \mathcal{E}(\mathcal{H})$, the space of trace class operators on $\mathcal{H}$. In our work $\mathcal{H}$ is the carrier space of a unitary representation of a Lie group, the symmetry group of the system.  

The quantum states in the standard theory are replaced by quantum conditions, the quantum conditions being open subsets of the state space $\EsubS(\mathcal{M})$ which has the weak topology generated by the functions $a_{Q}(\cdot)$ defined by $\hat A \in \mathcal M$, $a_{Q}(\rho) = Tr \rho \hat A, \; \forall \rho \in \EsubS(\mathcal{M})$.
If the quantum condition of the system is given by an open subset $U$ of the state space $\EsubS(\mathcal{M})$ then the attribute represented by $\hat A$ has the value $a_Q(U)$, $U$ giving the extent to which the quality $\hat A$ takes its qr-number value.

Topos theory is used in place of set theory because of its logic and real numbers. The logic of topos theory is \emph{intuitionistic}. It retains the law of non-contradiction and excludes the law of the excluded middle. For a spatial topos $\Shv (X)$, the extent to which a proposition is true is given by an open subset of the topological space $X$. The sheaf $\Omega$, defined for any open subset $W \subseteq X$ by $\Omega(W) = \{ U | U \subset W, U \in \mathcal{O}(X)\}$, is a subobject classifier for $\Shv(X)$. 

The topos $Shv(X)$ has an object called the sheaf of {\em Dedekind} reals $\RsubD(X)$. The local sections of $\RsubD(X)$ over open subsets $W$ of $X$ are its real numbers defined to extent $W$. $\RsubD(X)$  is equivalent to the sheaf $C(X)$ of continuous real-valued functions on $X$ \cite{maclane}. These topoidal real numbers are as mathematically acceptable as the standard real numbers of set theory which are equivalent to $\RsubD(X)$ when $X$ is the one point space. Which real numbers should used in a physical theory depends upon their fitness for the task of representing the numerical values of the physical qualities of the theory. 

In the qr-number interpretations \cite{adelman2},\cite{arxives},\cite{arxives2},\cite{durt2}, the topological space $X =  \EsubS(dU(\mathcal{E}(\mathcal{G})))$. The $O^{\ast}$-algebra comes from a unitary representation $\hat U$ of a Lie group $G$. The underlying $\ast$-algebra is the enveloping algebra $\mathcal{E}(\mathcal{G})$ of the Lie algebra $\mathcal{G}$ of $G$. In the theory of qr-numbers, the
 O$^{\ast}$-algebra $\mathcal{M}$ is the infinitesimal representation $dU$ of the enveloping algebra $\mathcal{E}(\mathcal{G})$ obtained from a unitary representation $U$ of $G$ on a separable Hilbert space $\mathcal{H}$. The properties of $dU(\mathcal{E}(\mathcal{G}))$ are used to determine the topological properties of the space $X = \EsubS(dU(\mathcal{E}(\mathcal{G})))$. 

The discussion of $O^{\ast}$-algebras and their state spaces relies heavily on the books of Smudgeon and Inoue, \cite{inoue}, although it contains some new results specific to the needs of qr-numbers. The discussion of topos theory in the book of MacLane and Moerdijk\cite{maclane} is required to fill out the general notion of Dedekind real numbers in a topos. The different constructions of real numbers in a topos $\Shv (X)$ of sheaves on a topological space $X$ are reviewed in Stout \cite{stout}. The brief introduction to these topics is supplemented in the Appendix which contains some useful definitions and results both from the theory of operators and operator algebras on Hilbert spaces: functional analysis on locally convex spaces, Lie groups, Lie algebras and enveloping Lie algebras,unitary  representations and $\ast$-representations of enveloping algebras, and topos theory: real numbers in a topos, sheaves, sections .

The expectation values of the standard quantum mechanical formalism are not qr-numbers but are order theoretical infinitesimal  qr-numbers, because there is no open set on which $x > 0 \lor x< 0$ holds. Each  infinitesimal qr-number is an \emph{intuitionistic nilsquare infinitesimal}\cite{jlbell} in the sense that it is not the case that it is not a nilsquare infinitesimal. This means that the standard quantum formulae provide infinitesimal approximations to qr-number formulae.
\subsection{Topos theory} 
In this paper only the example of a spatial topos, $Shv(X)$, the category of sheaves on the topological space $X$ will be discussed. 

A {\em spatial topos} is a category of sheaves on a topological space. The objects of this category are
sheaves over the topological space and the arrows are sheaf morphisms, that is, an arrow is a continuous function that maps a sheaf $Y$ to a sheaf $Y'$ in such a way that it sends fibres in $Y$ to fibres in $Y'$, equivalently, it sends sections of $Y$ over $U$ to sections of $Y'$ over $U$, where $U$ is an open subset of the base space.

 In $Shv(X)$ the subobject classifier is $\Omega = \mathcal O(X)$, the open subsets of $X$, therefore  the internal logic of $Shv(X)$ is intuitionistic and only is Boolean when $X$ is the one point set or when the topology on $X$ is $\{ \emptyset, X \}$. 

Mathematical arguments work in topos theory in much the same way as in set theory except that only constructive arguments are acceptable; neither the law of excluded middle nor the axiom of choice can be used. This is particularly straightforward in $Shv(X)$, in which sheaves play the role of sets, sub-sheaves substitute for subsets and local sections for the elements of a set. Then we can make proofs that look like proofs about sets. In the Kripke-Joyal semantics\cite{maclane} applied to $Shv(X)$, if a property is true for a sheaf $A$ then it is true for all subsheaves $A|_U$ for $U \in \mathcal O(X)$, if it true for each subsheaf  $A|_{U_{\alpha}}$, where $\{ U_{\alpha} \}_{\alpha \in J}$ form an open cover of $W \in \mathcal O(X)$ then it is true for $A|_W$, and there exists a local section $f|{W}$ for which the property is true means that there is an open cover $\{ W_{\alpha} \}_{\alpha \in J}$ of $W$ such that for each $\alpha$ there is a local section $f|{W_{\alpha}}$ for which the property holds.

\subsubsection {Real Numbers in Spatial Toposes}

{\em Dedekind numbers} are defined to be the completion of the rational numbers obtained by using cuts, and {\em Cauchy numbers} are defined as the completion of the rationals obtained by using Cauchy sequences. These different constructions can only be shown to be equivalent by using either the \AxChoice\  or the law of the excluded middle.\cite{maclane} Therefore, when intuitionistic
logic is assumed, these two types of real numbers are not equivalent.

It has been shown.\cite{maclane} that in a spatial topos the sheaf of rational numbers is the sheaf whose sections over an open set $U$ are given by locally constant functions from $U$ with values in $\QQ$ while the sheaf of Cauchy reals $\RsubC(X)$ is the sheaf whose sections over an open set $U$ are given by locally constant functions from $U$ with values in $\RR$. Here a function is locally constant if it is constant on each connected open subset of its domain.
 
On the other hand the sheaf of Dedekind reals  $\RsubD(X)$ is the sheaf whose sections over $U$ are given by continuous functions from $U$ to $\RR$. The Cauchy reals $\RsubC(X)$ form a proper sub-sheaf of the Dedekind reals unless the underlying topological space $X$ is the one point space.

The sheaf of Dedekind reals $\mathbb{R}_{\mathrm{D}}(X)$ satisfies, for proofs see \cite{maclane},\cite{stout} or \cite{johnstone},

1. It contains the integers $\mathbb{Z}(X)$ and rational numbers $\mathbb{Q}(X)$ (as sheaves of locally constant functions $ X \to \mathbb{Z}\; \text{or}\;\mathbb{Q}$). 

2. It has orders $<$ and  $\leq$ compatible with those on the rationals $\mathbb{Q}(X)$, but the order $<$ is partial not total because trichotomy, $x>0 \lor x=0 \lor x<0$, is not satisfied.

3. Algebraically it is a field; closed under the binary operations  +  and $\times$, has $0 \ne 1$
and if a number $b$ is not invertible then $b = 0$ (called a residue field by Johnstone \cite{johnstone}).  

4.Topologically it has a distance function $|\cdot|$ defining a metric on it. It is a complete metric space in which the rationals, $\mathbb{Q}(X)$, are dense. A number $b$ is apart from $0$ iff  $|b| > 0$. $\RsubD(X)$ is an apartness field, i.e., $\forall b \in \RsubD(X), \;  |b| > 0 $ iff $b$ is invertible.

\subsubsection { Dedekind real numbers for quantum systems}
The qr-numbers are the Dedekind reals $\RsubD(X)$ when $X = \EsubS(\mathcal{M})$.  $\EsubS(\mathcal{M})$ is  the space of quantum states of the physical system obtained as linear functionals on the $^{\ast}$-algebra of operators $\mathcal{M}$ that represent the physical qualities of the system.  The structure of the qr-numbers will be discussed in detail after the properties of the quantum state space $\EsubS(\mathcal{M})$ and the sheaf of locally linear functions  $\AAA(\EsubS(\mathcal M))$  have been exposed.
\subsection{O$^{\ast}$-algebras} 
Schm\"udgen \cite{inoue} is the basic reference for the mathematics of O$^{\ast}$-algebras.
O$^{\ast}$-algebras are attractive because they allow us to directly represent physical qualities such as energy, momentum, angular momentum and position.
   
If $\mathcal D$ is a dense subset of a
Hilbert space $\scrH$, let $\mathcal L(\mathcal D)$ (resp. 
$\mathcal L_{c}(\mathcal D)$) denote the 
set of all (resp. all closable) linear operators from $\mathcal D$ to $\mathcal D$, see $\S$\ref{ops}. 
If $\hat A^{\ast}$ is the Hilbert space adjoint of a linear operator $\hat A$ whose domain, $\text{dom}(\hat A)$, is $\mathcal{D}$ put
\begin{equation} 
\mathcal{L}^{\dagger}(\mathcal D) = \{ \hat A \in \mathcal L(\mathcal D) ; \text{dom}(\hat A^{\ast})
\supset \mathcal D,  \hat A^{\ast} \mathcal D \subset \mathcal D\}
\end{equation}
Then $\mathcal L^{\dagger}(\mathcal D) \subset \mathcal L_{c}(\mathcal D) \subset \mathcal L(\mathcal D) $ where $ \mathcal L(\mathcal D) $ is an algebra with the usual operations for linear operators with a common invariant domain: addition $\hat A +\hat  B$, scalar multiplication $\alpha \hat A$ and
non-commutative multiplication $\hat A\hat B$. Furthermore $\mathcal L^{\dagger}(\mathcal D)$  is a $\ast$-algebra with the involution $\hat A \to \hat A^{\dagger} :\equiv \hat A^{\ast} |_{\mathcal D}$. Note that if $\exists \hat A \in \mathcal L^{\dagger}(\mathcal D) $ that is closed, then $\mathcal D = \scrH$ and 
$\mathcal L^{\dagger}(\mathcal D) = \mathcal B(\scrH)$ the algebra of all bounded linear operators on $\scrH$.
\begin{definition}
An O-algebra on $\mathcal D \subset \scrH$ is a subalgebra of $\mathcal L(\mathcal D)$ that is contained in $\mathcal L_{c}(\mathcal D)$ whilst an O$^{\ast}$-algebra on $\mathcal D \subset \scrH$ is a $\ast$-subalgebra of $\mathcal L^{\dagger}(\mathcal D)$. 
\end{definition} 
Let $\mathcal M$ be an O-algebra on $\mathcal D \subset \scrH$, then the natural graph topology of $\mathcal M$ on $\mathcal D$, denoted $t_{\mathcal M}$, is the locally convex topology defined by the family of seminorms $\{ \| \cdot \|_{\hat A} ; \hat A \in \mathcal M \}$ where $\| \phi \|_{\hat A} =  \| \hat A\phi \|, \phi \in \mathcal D$. It is the weakest locally convex topology on $\mathcal D$ relative to which each operator $\hat A \in \mathcal M$
is a continuous mapping of $\mathcal D$ into itself. Every O$^{\ast}$-algebra $\mathcal M$  is a directed family, Schm\"udgen, Definition 2.2.4,\cite{inoue}, so that a mapping $\hat A$ of $\mathcal D$  into $\mathcal D$ is continuous iff for each semi-norm $\|\cdot\|_{\hat X}$ on  $\mathcal D$ there is a semi-norm $\|\cdot\|_{\hat Y}$ on  $\mathcal D$ and a positive real number $\kappa$  such that 
\begin{equation}
\|\hat A \phi  \|_{\hat X}\leq \kappa \|\ \phi \|_{\hat Y}
\end{equation}
for all $\phi \in \mathcal D$. Therefore any $\hat A \in \mathcal M$ defines a continuous map from $\mathcal D$ to $\mathcal D$ by taking $\hat Y = \hat X \hat A$ and $\kappa = 1$ for all $\hat X \in \mathcal M$. This shows that any $\hat A \in \mathcal M$ is bounded as a linear map from $\mathcal D$ to $\mathcal D$.

We will use the pre-compact topology on $\mathcal D$ which is determined by a directed family of semi-norms $p_{\scrK,\scrN } (\hat A) := \sup_{\phi \in \scrK}  \sup_{\psi \in \scrN} | \langle \hat A \phi, \psi\rangle | $ where  $\scrK, \scrN$ range over pre-compact subsets of $\mathcal D$ in the topology, $t_{\mathcal M}$, see Schmudgen\cite{inoue}, Section 5.3. If $\mathcal D$  is a complete locally convex Hausdorff space with respect to the topology  $t_{\mathcal M}$, then any subset whose closure is compact  is pre-compact. $\mathcal D$ is a Frechet space if it is metrizable, e.g., when the topology $t_{\mathcal M}$ is generated by a countable number of semi-norms.  

\subsubsection{The $O^{\ast}$-algebra $dU(\mathcal{E}(\mathcal{G}))$.}  
In this paper each $O^{\ast}$-algebra comes from a unitary representation $\hat U$ of a Lie group $G$ on a Hilbert space $\mathcal{H}$. Its $\ast$-algebra is the enveloping algebra $\mathcal{E}(\mathcal{G})$ of the Lie algebra $\mathcal{G}$ of $G$.
Given a unitary representation $U$ of $G$ on $\mathcal{H}$, a vector $\phi \in \mathcal{H}$ is a 
$C^{\infty}$-vector for $U$ if the map $g \to U(g)\phi$ of the $C^{\infty} $ manifold $G$ into $\mathcal{H}$ is a $C^{\infty}$-map. The set of $C^{\infty}$-vectors for $U$ is denoted $\mathcal{D}^{\infty}(U)$, it is a dense linear subspace of $\mathcal H$ which is invariant under $\hat U(g), \; g \in G$. 

The representation of the Lie algebra $\mathcal{G}$ of $G$ is obtained from the unitary representation $U$ of $G$ by defining $\forall x \in \mathcal{G}$ the operator $dU(x)$ with domain $\mathcal{D}^{\infty}(U)$ as
\begin{equation}
dU(x)\phi = \frac{d}{dt}U(\exp tx)\phi|_{t=0} = \lim_{t\to0} t^{-1}(U(\exp tx ) - I)\phi, \; \phi \in \mathcal{D}^{\infty}(U)
\end{equation}

$dU(x)$ belongs to a $\ast$-representation of $\mathcal{G}$ on $\mathcal{D}^{\infty}(U)$ which has a unique extension to a $\ast$-representation of $\mathcal{E}(\mathcal{G})$ on $\mathcal{D}^{\infty}(U)$ called the \emph{infinitesimal representation} of the unitary representation $U$ of $G$. Each operator $\imath  dU(x)$ is essentially self-adjoint on $\mathcal{D}^{\infty}(U)$.
The graph topology $\tau_{dU}$  is generated by the  family of semi-norms  $\| \cdot \|_{dU(y)} $, for $y \in \mathcal{E}(\mathcal{G})$. Then $\mathcal{D}^{\infty}(U) $ is a Frechet space when equipped with the graph topology $\tau_{dU}$. 

The graph topology $\tau_{dU}$ on $\mathcal{D}^{\infty}(U)$ can be generated by other families of semi-norms. 
\begin{lemma}\cite{inoue} Schm\"udgen, Corollary 10.2.4.
Let a be an elliptic element of $ \mathcal{E}(\mathcal{G}) $, then $\mathcal{D}^{\infty}(U) = \mathcal{D}^{\infty}(\overline{dU(a)})$ and the graph topology $\tau_{dU}$ on $\mathcal{D}^{\infty}(U) $ is generated by the family of semi-norms $\| \cdot \|_{dU(a)^{n}}, \; n \in \NN_{0}$.
\end{lemma}
 
The two families of semi-norms $\{ \| \cdot \|_{dU(a)^{n}}, \; n \in \NN_{0}\}$ and $\{\| \cdot \|_{dU(y)}, \; y \in  \mathcal{E}(\mathcal{G}) \}$ are equivalent because they both generate the graph topology on $\mathcal{D}^{\infty}(U)$ and the first is clearly a directed family so for each $y \in  \mathcal{E}(\mathcal{G})$ there exists a constant $K$ and an integer $n\in \NN$ so that for all $\phi \in \mathcal{D}^{\infty}(U)$,
\begin{equation}
\|dU(y)\phi\| \leq K \|dU(a)^{n}\phi \|.
\end{equation}

\subsection{The state spaces  \EsubS($\mathcal M$)\ } 

Let the $\ast$-algebra  $\mathcal A$ of physical qualities  be represented by the O$^{\ast}$-algebra 
$\mathcal M$ defined on the dense subset $\mathcal{D} \subset \scrH$. We assume that $\mathcal M$ has a unit element, the identity operator $\hat I$.

A linear functional $f$ on the O$^{\ast}$-algebra $\mathcal M$ is a linear map from $\mathcal M$ 
to the standard complex numbers $\mathbb C$, it is strongly positive iff 
$f(\hat X)\geq{0}$ for all $\hat X\geq{0}$ in $\mathcal M$. 

\begin{definition}
The states on $\mathcal M $ are the strongly positive 
linear functionals on $\mathcal M$ that are normalised to take the 
value 1 on the unit element $\hat I $ of $\mathcal M $. The state space $\EsubS (\mathcal M)$ of the O$^{\ast}$-algebra $\mathcal M$ is the set of all states on $\mathcal M$.
\end{definition}
\begin{definition}
A bounded operator $\hat B$ on $\scrH$ is trace class iff $Tr |\hat B| < \infty$.
A trace functional on $\mathcal M$ is a functional of the form $ \hat A \in \mathcal M \mapsto Tr (\hat B \hat A) $ for some trace class operator $\hat B$.
\end{definition}
The states of the O$^{\ast}$-algebra $\mathcal M$ are given by trace class operators.
\begin{theorem}\cite{inoue}\label{TH1}
Every strongly positive linear
functional  on $\mathcal M $ is given by a trace functional.
\end{theorem}

The state space $\EsubS(\mathcal M)$ is contained in the convex
hull of projections $\scrP$ onto one-dimensional subspaces
spanned by unit vectors $\phi  \in \mathcal D$. If 
$\hat \rho \in \EsubS (\mathcal M)$ then there is an orthonormal set of vectors 
$\{\phi_{n}\}_{n\in \NN'}$  in $\mathcal D$  such that
$\hat \rho = \sum_{n \in \NN'}\lambda_{n}
|\phi_{n}\rangle\langle\phi_{n}| = 
\sum_{n \in \NN'}  \lambda_{n} \scrP_n $ where
$\scrP_n = |\phi_{n}\rangle\langle\phi_{n}|$ is the orthogonal projection onto the
one-dimensional subspace spanned by $\phi_{n}$,
$\lambda_{n} \in \RR, 0 \leq \lambda_{n} \leq 1 $,$\sum_{n \in \NN'}
 \lambda_{n} = 1$ and $\NN' = \{ n \in \NN; \lambda_{n} \neq 0 \}$.  All states
in $\EsubS(\mathcal M)$ must satisfy the further  condition that as $n$ approaches
infinity the sequence $\{\lambda_{n}\}$ converges to zero faster than any
power of $1/n$ \cite{adelman2}. 

The following well-known example will be our guide. For the $C^{\ast}$-algebra $ \scrB( \scrH)$, the state space $\scrE = \EsubS(\scrB( \scrH))$ is composed of operators in $\scrT_{1}(\scrH)$ that are self-adjoint and normalised. The collection $\scrT_{1}(\scrH)$ of all trace class operators on $\scrH$ is a linear space over $\scrC$, it is a Banach space when equipped with the trace norm  $\nu(T) = Tr |T|$, where $|T| = \surd (T^{\ast} T)$. The open subsets of $\scrE$ in the trace norm topology are denoted  
$\nu(\hat \rho_{1} ; \delta) = \{ \hat \rho \; :Tr | \hat \rho - \hat \rho_{1}| < \delta\}$. $\scrE$ is compact in the weak$^\ast$ topology, the weakest topology on  $\scrE$ that makes continuous all the functionals $\hat \rho \to Tr (\hat \rho\hat B) , \hat B \in  \scrB( \scrH)$. Its sub-basic open sets are $ \scrN(\hat \rho_{0}  ; \hat B ; \epsilon) = \{\hat
\rho \; : |Tr \hat \rho \hat B - Tr \hat \rho_{0} \hat B| < \epsilon \}$ with $\hat B \in \mathcal B(\mathcal H)$ and $\epsilon > 0$.

In order to carry out a similar analysis for the states on an O$^{\ast}$-algebra $\mathcal M$, we define a subset $\mathcal{T}_{1}(\mathcal M)$ of $\scrT_{1}(\scrH)$. Let $\mathcal{T}_{1}(\mathcal M) \subset \scrT_{1}(\scrH)$ be composed of trace class operators $\hat T$ that satisfy $ \hat T\scrH \subset \mathcal D , \hat T^{\ast}\scrH \subset \mathcal D $ and $\hat A \hat T, \hat A \hat T^{\ast} \in
\mathcal{T}_{1}(\scrH)$ for all $\hat A \in \mathcal M $, that is,
$\mathcal{T}_{1}(\mathcal M)$ is a $\ast$-subalgebra of $\mathcal M$ satisfying $\mathcal M
\mathcal{T}_{1}(\mathcal M) = \mathcal{T}_{1}(\mathcal M)$\cite{inoue}. 
\begin{definition}
The state space $\EsubS(\mathcal M)$ for the $O^{\ast}$-algebra $\mathcal M$ is the set of normalized, self-adjoint operators in $\mathcal{T}_{1}(\mathcal M)$.
\end{definition}
$\EsubS(\mathcal M)$ has the weak topology generated by the functions $ a_{Q}(\cdot)$ where, given $\hat A \in  \mathcal M,  a_{Q}( \hat \rho) = Tr \hat A \hat \rho,  \forall  \hat \rho  \in \EsubS(\mathcal M) $. This topology is the weakest that makes all the functions $ a_{Q}(\cdot)$ continuous.
With parameters $\hat A \in \mathcal M, \epsilon > 0$ and $\rho_0 \in \EsubS(\mathcal M)$, the sets  $ \scrN(\hat \rho_{0}  ; \hat A ; \epsilon) = \{\hat
\rho \ ;|Tr \hat \rho \hat A - Tr \hat \rho_{0} \hat A| < \epsilon \}$  form an open 
sub-base for the weak topology on $\EsubS(\mathcal M)$. 
The weak topology on $\EsubS(\mathcal M)$ is stronger than the weak$^\ast$ topology on $\scrE$ restricted to $\EsubS(\mathcal M)$ because for any $\hat B \in \scrB( \scrH)$ the weak$^\ast$ sub-basic open set  $ \scrN(\hat \rho_{0}  ; \hat B ; \epsilon)$ is also open in the weak topology\cite{james}, \cite{reed}. 

It is also well known that for all $\rho_1 \in \EsubS(\mathcal M)$ and every $\delta > 0$, if $\hat A \in \scrB(\scrH)$ then, as subsets of $\EsubS(\mathcal M)$, $\nu(\hat \rho_{1} ; \delta) \subset  \scrN(\hat \rho_{1}  ; \hat A ;  K \delta) $ where $K = \| A \| $. The following generalizes this result to open subsets of labeled by essentially self-adjoint operators $\hat A \in \mathcal M$. It uses the precompact topology on $ \mathcal M$ determined by the family of semi-norms $ \mathcal P_{ M, N} (\hat A) = \sup_{\xi \in M, \eta \in N}
| \langle \hat A \xi, \eta \rangle | $ where $\{ M, N \}$ range over subsets of $\mathcal D$ precompact in  the graph topology $t_{\mathcal M}$, Schm\"udgeon,\cite{inoue}, Sections 5.3. A subset of $\mathcal D$ is precompact in the graph topology  $t_{\mathcal M}$ if its closure is compact in $t_{\mathcal M}$. 

\begin {lemma} If $\mathcal D$ is a Frechet space with respect to the topology $t_{\mathcal M}$, then 
for every essentially self-adjoint operator $\hat A \in \mathcal M $ and for every self-adjoint  
$\hat T \in \mathcal{T}_{1}(\mathcal M)$, 
\begin{equation}
 |Tr \hat A \hat T| \leq   p_{\hat T}(\hat A) \nu (\hat T).
\end{equation} 
$\mathcal P_{\hat T} (\hat A) = \sup_{\zeta_{n}} |\langle \hat A \zeta_{n},\zeta_{n} \rangle|  \leq \sup_{\zeta_{n}} \| \hat A \zeta_{n} \|$, where the set $\{\zeta_{n} \in \mathcal{D}\}$ is an orthonormal set of eigenvectors of $|\hat T|$ for eigenvalues $\lambda_{n} > 0 $.  If an eigenvalue $\lambda_{n}$ has multiplicity $s$, take an orthonormal basis $\{\zeta_{m}\}_{m=1}^{s}$ in its eigenspace as part of the set.
\end{lemma}  
 
$G_{\hat A}(\hat T) = Tr \hat A \hat T$ defines a linear functional $G_{\hat A}$ on $\mathcal{T}_{1}(\mathcal M)$. Note that $p_{\hat T}(\hat A)$ depends on $\hat T$, if it was independent then the inequality would show that  $G_{\hat A}$ was continuous with respect to the trace norm topology on $ \mathcal{T}_{1}(\mathcal M)$. 
  
\begin{proof}
Since $\mathcal D$ with the topology $t_{\mathcal M}$ is metrizable, a result of Grothendieck, \cite{kothe}, shows that the operator $\hat T\in \mathcal{T}_{1}(\mathcal M)$ has a canonical representation $\hat T = \sum _{n} \lambda_{n}  \langle \;, \zeta_{n} \rangle \eta_{n}$, where $\sum _{n}| \lambda_{n} | < \infty$ and the sequences $( \zeta_{n} )$ and $( \eta_{n} )$ converge to zero in $\mathcal D$. 
$\sum _{n} \lambda_{n}  \langle \;, \zeta_{n} \rangle \eta_{n}$ converges absolutely with respect to $\mathcal M $, that is, $\sum _{n} \lambda_{n}  \| \hat A   \eta_{n}\| \|\hat C \zeta_{n}\| < \infty$ for all $\hat A, \hat C \in \mathcal{M}$. 

If $\hat T$ is self-adjoint it has a canonical representation $\sum _{n} \lambda_{n}\langle \;,  \zeta_{n} \rangle \zeta_{n}$, the $\{\zeta_{n}\}$ being an orthonormal set of vectors in $\mathcal{D}$ where $\zeta_{n}$ is an eigenvector of $|\hat T|$ for the eigenvalue $\lambda_{n} > 0 $. By the spectral theorem, the series $\Sigma _{n} \lambda_{n}  \langle \;,\zeta_{n} \rangle \zeta_{n}$ is strongly convergent in $\scrH$ and $\Sigma_{n} | \lambda_{n}| \|\hat A \zeta_{n}\|^{2} < \infty$ for any 
$\hat A \in M $\cite{inoue}, Schm\"udgen,Lemma 5.1.10. Therefore the set $\{ \zeta_{n} \}$ has at most the single limit point $0 \in \mathcal D$ in the graph topology.
  
For a fixed $\hat T \in \mathcal{T}_{1}(\mathcal M)$ and any $\hat A \in M $,
\begin{equation}
 |Tr \hat A \hat T | = | \Sigma _{n} \langle (\hat A \lambda_{n} \zeta_{n}), \zeta_{n} \rangle | \leq (\Sigma_{n} |\lambda_{n}|) \mathcal P_{ \hat T} (\hat A).
\end{equation}
The set $V(\hat T) =  \{ \zeta_{n} \}$ of eigenvectors of $|\hat T|$ form a pre-compact set in $\mathcal D (t_{\mathcal M})$ so that  $\mathcal P_{\hat T} (\hat A) = \sup_{\zeta_{n}} |\langle \hat A \zeta_{n},\zeta_{n} \rangle| \leq \sup_{\zeta_{n}} \| \hat A \zeta_{n} \|$ because $\| \zeta_{n}\| = 1$ for all $n$. $\nu (\hat T) = (\Sigma_{n} |\lambda_{n}|) $ is the trace norm of $\hat T$.  We cannot immediately infer that $G_{\hat A}$ is continuous with respect the trace norm restricted to $\mathcal{T}_{1}(\mathcal M)$ because $ \mathcal P_{\hat T} (\hat A)$ depends on $\hat T$. If the supremum was taken over all of $\mathcal{D}$ the argument could be extended to a proof of the desired continuity of  $G_{\hat A}$.
\end{proof}
There are two classes of operators $\hat A$ for which we can prove continuity of $G_{\hat A}$ with respect the trace norm restricted to $\mathcal{T}_{1}(\mathcal M)$. In the first, the self-adjoint operator $\hat A$ is bounded on $\mathcal{H}$ then $ \mathcal P_{ \hat T} (\hat A) \leq \sup_{\zeta_{n}} \| \hat A \zeta_{n} \| \leq  \|\hat A \|$. The second contains essentially self-adjoint operators $\hat A \in \mathcal{M}$ which are `relatively bounded' with respect to a positive e.s-a. operator $\hat N \in \mathcal{M} $ with $\hat N + \hat I$ an invertible operator that maps $\mathcal{D}$ into itself. Relative bounded means that  $\hat A (\hat N + \hat I)^{-1}$ is a bounded operator. Then $ \mathcal P_{ \hat T} (\hat A) \leq
\sup_{\zeta_{n}} \| \hat A \zeta_{n} \| \leq \sup_{\phi \in \mathcal{D}} \| \hat A \phi \| \leq \sup_{\psi \in \mathcal{D}} \| \hat A (\hat N + \hat I)^{-1}\psi \| \leq  \|\hat A (\hat N + \hat I)^{-1}\|$. 
\subsection{When $\mathcal{M}$ is  $dU(\mathcal{E}(\mathcal{G}))$}
If the O$^{\ast}$-algebra $\mathcal{M}$ is the infinitesimal representation $dU$ of the enveloping algebra $\mathcal{E}(\mathcal{G})$ obtained from a unitary representation $U$ of a Lie group $G$, then the required positive e.s-a. operator $\hat N$ is $dU(  - \Delta)$ where $\Delta = \sum_{i=1}^{d} x_{i}^{2}$ is the Nelson Laplacian in the enveloping algebra of the Lie algebra $\mathcal{G}$ with basis $\{ x_{1},x_{2},......,x_{d}\}$.

To prove this let $U$ be a unitary representation of a Lie group $G$ on a Hilbert space $\mathcal{H}$ with $dU$ the infinitesimal representation of its Lie algebra $\mathcal{G}$ on the set $\mathcal{D}^{\infty}(U)$ of $C^{\infty}$-vectors. $dU$ is also a $\ast$- representation of the enveloping algebra $\mathcal{E}(\mathcal{G})$ on $\mathcal{D}^{\infty}(U)$ which is dense in $\mathcal{H}$.
\begin{theorem} \label{TH2}
Each essentially self-adjoint (e.s-a.) operator $\hat A = dU(x)$ that represents an element $x \in \mathcal{E}(\mathcal{G}) $ defines a linear functional $G_{\hat A}$ on 
$\mathcal{T}_{1}(dU(\mathcal{E}(\mathcal{G}))$ that is continuous with respect to the trace norm topology on $ \mathcal{T}_{1}(dU(\mathcal{E}(\mathcal{G}))$. 
\end{theorem}
\begin{proof} 
Write $\mathcal M$ for $dU(\mathcal{E}(\mathcal{G}))$.
From the preceding lemma we have that for a fixed $\hat T \in \mathcal{T}_{1}(\mathcal M)$ and any $\hat A \in M $,
\begin{equation}
 |Tr \hat A \hat T | \leq (\Sigma_{n} |\lambda_{n}|) \mathcal P_{ \hat T} (\hat A) \leq \nu (\hat T)\sup_{\phi \in \mathcal{D}^{\infty}(U)} \| \hat A \phi \|.
\end{equation}
Now Lemma 3 shows that the graph topology on $\mathcal{D}^{\infty}(U)$ is generated by the families of semi-norms $\{ \|\cdot\|_{dU(x)} = \|dU(x)\cdot\|\; ; x \in \mathcal{E}(\mathcal{G}) \}$ and $\{ \| \cdot \|_{dU(1 - \Delta)^{n}}, \; n \in \NN_{0} \}$ with the consequence that for each $y \in \mathcal{E}(\mathcal{G})$ there exists an integer $m \in \NN$ and a constant $K$ with
\begin{equation}
\|dU(y)\phi\| \leq K \|dU((1- \Delta)^{m}) \phi \|
\end{equation}
for all $\phi \in \mathcal{D}^{\infty}(U)$, where $\mathcal{D}^{\infty}(U) =  \mathcal{D}^{\infty}(\overline{dU(1 - \Delta)})$, and $\mathcal{D}^{\infty}(\overline{dU(1 - \Delta)}) = \bigcap_{n\in\NN} \mathcal{D}(\overline{dU(1 - \Delta)}^{n})$.
For any integer $m \in \NN$ the operator $dU((1- \Delta)^{m})= (dU(1- \Delta))^{m}$ is a positive, essentially self-adjoint operator on $\mathcal{D}^{\infty}(U) $ that maps  $\mathcal{D}^{\infty}(U) $ onto itself, so that for each $\phi \in \mathcal{D}^{\infty}(U)$ there exists $\psi \in \mathcal{D}^{\infty}(U)$ such that $\phi = dU((1- \Delta)^{m})^{-1} \psi$. This implies that for each $y \in \mathcal{E}(\mathcal{G})$ equation (8) becomes,
\begin{equation}
\|dU(y)\phi\| = \|dU(y)dU((1- \Delta)^{m})^{-1}  \psi\| \leq K \|\psi\|.
\end{equation}
Therefore on putting $\hat A = dU(y)$ and normalizing $\psi$, the inequality in equation (7) becomes
\begin{equation}
 |Tr \hat A \hat T | \leq \nu (\hat T)\sup_{\psi \in \mathcal{D}^{\infty}(U)} \| \hat A dU((1- \Delta)^{m})^{-1} \psi \|/ \|\psi\| \leq \nu (\hat T)  K .
\end{equation}
 
Let $\hat \rho_1, \hat \rho \in \EsubS(\mathcal{E}(\mathcal{G}))$ and put $ \hat T = (\hat \rho -  \hat \rho_1)$ and $K = K(\hat A)$, which is independent of $\hat T$, then 
\begin{equation}
|Tr (\hat A \hat \rho - \hat A \hat \rho_1)| \leq  K(\hat A) Tr | \hat \rho - \hat \rho_1|.
\end{equation} 
\end{proof}
\begin{corollary}\label{CR7}
Each function $a_{Q}(\cdot)$ for an essentially self-adjoint operator $\hat A \in dU(\mathcal{E}(\mathcal{G}))$  is continuous from $\EsubS(dU(\mathcal{E}(\mathcal{G})))$ to $\RR$ in the trace norm topology restricted to  $\EsubS(dU(\mathcal{E}(\mathcal{G})))$. 
\end{corollary}
\begin {corollary}\label{CR8}
The open sets  $\{\nu (\hat \rho_{1} ; \delta); \hat \rho_{1} \in \EsubS(dU(\mathcal{E}(\mathcal{G}))), \delta > 0 \}$ form an open basis for the weak topology on $\EsubS(dU(\mathcal{E}(\mathcal{G})))$.
\end{corollary}
\begin{proof}
Let
$\hat \rho \in \scrN(\hat \rho_{0} ; \hat A ; \epsilon)$ and suppose
that $|Tr \hat \rho \hat A - Tr \hat \rho_{0} \hat A | = \mu < \epsilon$.
Now take $(\epsilon - \mu)/2 = \epsilon ' < \epsilon$ and choose 
$\delta' > 0$ such that $|Tr \hat \rho \hat A - Tr \hat \rho_{1} \hat A|
< \epsilon'$ when $Tr | \hat \rho - \hat \rho_{1} | < \delta' $.

 Then $\nu (\hat \rho ; \delta ') \subset  \scrN(\hat \rho_{0} ;
\hat A ; \epsilon)$ because $|Tr \hat \rho_{1} \hat A - Tr \hat \rho_{0}
\hat A | < |Tr \hat \rho_{1} \hat A - Tr \hat \rho \hat A| + |Tr \hat
\rho \hat A - Tr \hat \rho_{0} \hat A | < \epsilon' + \mu = (\epsilon -
\mu)/2 + \mu = (\epsilon + \mu)/2 < \epsilon$. Therefore centred on each $\hat
\rho \in \scrN(\hat \rho_{0} ; \hat A ; \epsilon)$ there is an open set
$\nu(\hat \rho ; \delta ')$ contained in 
$\scrN(\hat \rho_{0} ; \hat A ; \epsilon)$.
Therefore  for each $\hat \rho \in \bigcap_{k=1,..n}\scrN(\hat \rho_{0}
; \hat A_k ; \epsilon_k)$, we can find $\delta_{k}' > 0$ such that 
$\
u(\hat \rho ; \delta_{k}') \subset \scrN(\hat \rho_{0}
; \hat A_k ; \epsilon_k)$ for each $k=1,..n$.
Take $\delta' = \min_{k=1,..n} \{\delta_{k}'\}$ then $\
u(\hat \rho ;
 \delta') \subset \bigcap_{k=1,..n}\scrN(\hat \rho_{0}
; \hat A_k ; \epsilon_k)$. But any open set in the weak topology
generated by the functions $a_Q$ is of the form $G = \bigcap_{k=1,..n}\scrN(\hat \rho_{0} ; \hat A_k ; \epsilon_k)$, therefore the $\nu(\hat \rho_{1} ; \delta)$ form an open base for  this topology.
\end{proof}
The boundaries of the open sets in the weak topology on $ \EsubS(dU(\mathcal{E}(\mathcal{G})))$ are composed of states $\hat \rho$ satisfying equations of the form $Tr \hat \rho \hat A = \alpha$, where $\alpha$ is a standard real number and $\hat A$ is a non-zero e.s-a.operator in the algebra $dU(\mathcal{E}(\mathcal{G}))$. The next results determine the interior of sets of states that satisfy such equations.

\begin{lemma}\label{LM9}
Given $\hat \rho \in  \EsubS (dU(\mathcal{E}(\mathcal{G})))$ with $Tr \hat \rho \hat A  = 0$ then for all $\epsilon > 0$ there exists  $\hat \sigma \in  \EsubS(dU(\mathcal{E}(\mathcal{G})))$ such that $\hat \sigma \in \nu(\hat \rho ; \epsilon)$ and $Tr \hat \sigma \hat A \ne 0$.
\end{lemma} 
\begin{proof} Write $\hat \rho = \sum_{i\in I} \lambda_{i} \hat P_{i}$ where $\forall i \in I, \lambda_{i} > 0$ and $\sum_{i\in I} \lambda_{i} = 1$ and where $\{\hat P_{i}\}_{i\in I}$ is an ortho-normal family of one dimensional projection operators. Given $\epsilon > 0$, consider three cases;
 
(i) if $Tr \hat P_{k} \hat A  = 0$ for a particular integer $k \in I$ and $Tr \hat P_{j} \hat A  \neq 0$ for a different integer $j \in I$ then choose a standard real number $0 < \alpha < \epsilon/2$ such that $\lambda_{k} > \alpha$ and set 
\begin{equation}
\hat \sigma = (\lambda_{j} + \alpha) \hat P_{j} + (\lambda_{k} - \alpha) \hat P_{k} +  \sum_{i \neq j,k} \lambda_{i} \hat P_{i}
\end{equation}
then $\hat \sigma \in \EsubS(dU(\mathcal{E}(\mathcal{G})))$ and $Tr |\hat \sigma - \hat \rho | = 2\alpha < \epsilon$ with $Tr \hat \sigma \hat A = \alpha Tr\hat P_{j} \hat A \neq 0$.

(ii) if  $Tr \hat P_{i} \hat A  = 0$ for all $i \in I$ then the family $\{\hat P_{i}\}_{i\in I}$ cannot span the Hilbert space $\mathcal{H}$ unless $\hat A =0$ the zero operator. Let $\{\hat P_{k}\}_{k\in K}, K \subset I$ be the sub-family for which $\forall k \in K, Tr \hat P_{k} \hat A  = 0$ and choose a one dimensional projection operator $\hat Q$ such that $\forall k \in K, \hat P_{k} \perp \hat Q$ and $Tr \hat Q \hat A  \neq 0$. Choose  an operator $\hat P_{l}$ from $\{\hat P_{k}\}_{k\in K}$ and a standard real number $\alpha < \lambda_{l}$ with $\alpha < \epsilon$. Set  
\begin{equation}
\hat \sigma = \hat \rho  -   {\alpha\over 2} \hat P_{l}  +  {\alpha\over 2} \hat Q.
\end{equation}
Then $\hat \sigma \in \EsubS(dU(\mathcal{E}(\mathcal{G})))$ and $Tr |\hat \sigma - \hat \rho | = \alpha < \epsilon$ with $Tr \hat \sigma \hat A =  {\alpha\over 2} Tr\hat Q \hat A \neq 0$.

(iii) if $Tr \hat P_{i} \hat A  \neq 0$ for all $i \in I$ then there must be projection operators $\hat P_{j}, \hat P_{k}$ in the family $\{\hat P_{i}\}_{i\in I}$ with $Tr \hat P_{k} \hat A  \neq Tr \hat P_{j} \hat A $. Choose a standard real number $\alpha$ with $0 < \alpha < \min \{ {\epsilon\over 2}, \lambda_{j}\}$ and set 
\begin{equation}
\hat \sigma = (\lambda_{j} + \alpha) \hat P_{j} + (\lambda_{k} - \alpha) \hat P_{k} +  \sum_{i \neq j,k} \lambda_{i} \hat P_{i}
\end{equation}
Then $\hat \sigma \in \EsubS(dU(\mathcal{E}(\mathcal{G})))$ and $Tr |\hat \sigma - \hat \rho | = 2\alpha < \epsilon$ with $Tr \hat \sigma \hat A =  \alpha (Tr\hat P_{j} \hat A -  Tr\hat P_{k} \hat A) \neq 0$.
\end{proof}
\begin{proposition} \label{PR10}
Let $\hat A \in dU(\mathcal{E}(\mathcal{G}))$ be a non-zero essentially self-adjoint operator, then
 $\text{int}\{\hat \rho | Tr \hat \rho \hat A = 0 \} = \emptyset$ where the interior is taken with respect to the weak topology on $\EsubS(dU(\mathcal{E}(\mathcal{G})))$.
\end{proposition}
\begin{proof} Let $\text{int}_t$ be the interior operator with respect to the trace norm topology on $\EsubS(dU(\mathcal{E}(\mathcal{G})))$. Lemma 9 implies that 
$\text{int}_t\{\hat \rho | Tr \hat \rho \hat A = 0 \} = \emptyset$
but every open set in the weak topology is a union of open sets in the trace norm topology, therefore $\text{int}\{\hat \rho | Tr \hat \rho \hat A = 0 \}  \subseteq \text{int}_t\{\hat \rho | Tr \hat \rho \hat A = 0 \}$ because the interior of a set is the union of all its open subsets.
\end{proof}
\begin{corollary}\label{CR11}
For any standard real number $\alpha$, $\text{int}\{\hat \rho | Tr \hat \rho \hat A = \alpha \} = \emptyset$.
\end{corollary}
Obviously, each state $\hat \rho \in \EsubS(dU(\mathcal{E}(\mathcal{G})))$ has empty interior, $\text{int}\{ \hat \rho \}  = \emptyset$ and this remains true for a disjoint countable union of individual states,  $\text{int}\{ \coprod_{\hat \rho} \hat \rho \} = \emptyset$.

\subsection { Quantum real numbers}
 The qr-numbers of a system whose are the Dedekind reals $\RsubD(X)$ when $X = \EsubS(\mathcal{M})$. 
By the construction of the topology on $\EsubS(\mathcal M)$,
for any essentially self-adjoint operator $A \in \mathcal M$  the function
 $a_Q(\rho)= Tr \hat \rho \hat A$  is a globally defined continuous
function and therefore determines a global section of $C(\EsubS(\mathcal M))$.  
The functions $a_Q(W)$, with 
domains given by open subsets $\emptyset \neq W \subset  \EsubS(\mathcal M)$, are interpreted as the numerical values of the physical quality that is represented by the operator $\hat A \in \mathcal M$.  Real numbers of this form are a proper sub-sheaf $\AAA(\EsubS(\mathcal M)) $ of $C(\EsubS(\mathcal M))$ called the sheaf of \emph{locally linear functions}.\cite{adelman2} Every qr-number is a continuous function of locally linear functions in $\AAA(\EsubS(\mathcal M))$\cite{adelman2}.

For each open set, $\emptyset \neq W \subset \EsubS(\mathcal{A})$, the qr-numbers $C(\EsubS(W))$, defined to extent $W$, have the same properties as the globally defined qr-numbers $C(\EsubS(\mathcal M))$. Therefore we restrict our discussion to the latter.

\subsubsection{The ontological and epistemological conditions}
\begin{definition} The ontological conditions of the system whose qualities are represented by operators in $dU(\mathcal{E}(\mathcal{G})))$ are given by the open sets $W$ in the weak topology on $ \EsubS(dU(\mathcal{E}(\mathcal{G})))$.
For any essentially self-adjoint (e.s-a.) operator $\hat A \in dU(\mathcal{E}(\mathcal{G})) $ the qr-number value of the quality represented by $\hat A$ in the ontological condition $W$ is $a_{Q}(W)$.
\end{definition}

We will distinguish between the ontological and epistemological condition of a quantum system. 
If a system is prepared in an open set $W$ of state space, then the {\em epistemological condition} of the system is a sieve $S(W)$ generated by $W$. A sieve $S(W)$ on an open set $W$ is a family of open subsets of $W$ with the property if $U\in S(W)$ and $V \subset U$ then $V \in S(W)$\cite{maclane}. We will usually take $S(W)$ to be the family of all non-empty open subsets of $W$ because for any open set $V \subset  W$, the values of qualities defined to extent $V$ will satisfy the experimental restrictions imposed on qualities defined to extent $W$. The ontological value of $\hat A$ could be $a_{Q}(V)$ for any non-empty $V \subset  W$.

There is an important difference between the average value of a physical quantity
represented by $\hat A$  at a particular pure state
$\hat \rho_0$ and the qr-number $a_Q(W_0)$ defined by $\hat A$ on an ontological condition  $W_0$.

We will write $\langle\hat A\rangle_{\hat \rho_0}$ for the average value 
of $\hat A$ in the state $\hat \rho_0$,  $\langle\hat A\rangle_{\hat
\rho_0} = Tr \hat \rho_0 \cdot \hat A$.
 If $\langle\hat A\rangle_{\hat \rho_0} = \langle\hat B\rangle_{\hat 
\rho_0}$ then it doesn't follow that $\hat A  =  \hat B$ as operators.
However, for any non-empty subset $W_0$ the qr-numbers $a_Q(W_0)
= b_Q(W_0)$ if and only if the defining operators are equal, $\hat A =
\hat B$.

\begin{proposition} \label{PR13}
Let $ W_0$ be an open subset of $\EsubS(dU(\mathcal{E}(\mathcal{G})))$ and suppose that $a_Q(W_0) = 0$. If $ W_0 \neq \emptyset$ then $\hat A = \hat O$, the zero operator.
\end{proposition}
\begin{proof} By Proposition 9 if $\hat A$ is not the zero operator then the interior of $W_{0}$ must be empty, but the interior of $W_{0}$ is $W_{0} \neq \emptyset$ by hypothesis, so that $\hat A = \hat O$ the zero operator.
\end{proof}

\begin{corollary}\label{CR14}
Let $W$ be a non-empty open subset  of $\EsubS(dU(\mathcal{E}(\mathcal{G})))$, if
$a_Q(W) = k$, a standard real number, then $\hat A = k \hat I$. That is
if a qr-number is equal to a standard real number to the extent
$W$, then it equals the standard real number to full extent.
\end{corollary} 

\begin{corollary} \label{CR15}
On any non-empty open set $W$, two qr-numbers are equal, $a_Q(W)
= b_Q(W)$, if and only if the defining e.s-a. operators are equal, $\hat A = \hat
B$. Therefore knowing $a_Q(W)$ is equivalent to knowing the operator $\hat
A$. \end{corollary}
For any non-empty open set $W_0$, the qr-number $a_Q(W_0)$ provides more information than the mean value $\langle\hat A\rangle_{\hat \rho_0}$ at any $\hat \rho_0 \in W_0$.

\begin{corollary} \label{CR16}
Let $W$ be any non-empty open set.
 If we know the one parameter family of qr-numbers
$a(t)_Q(W)$ for all standard real numbers $t$ belonging to some
interval $J$ then we know the one-parameter family of operators $\hat
A(t)$ with $t \in J$ such that 
$a(t)_Q(W)$ is defined by the function $a(t)_Q(\hat \rho) = Tr \hat
\rho \hat A(t)$ for  $\hat \rho \in W$.
\end{corollary}

\subsubsection{Properties of locally linear quantum real numbers}
We list some properties of $\AAA(\EsubS(\mathcal M))$ whose sections, the functions $a_Q$, are \emph{locally linear quantum real numbers}.
\begin{definition}\cite{adelman1}
Let $U$ be an open subset of $\EsubS(\mathcal M)$.  A  function $f : U \to \RR$ is {\em locally linear} if each $\rho \in U$ has an open neighborhood $U_{\rho} \subset U$ with an essentially self-adjoint (e.s-a.) operator  $\hat A \in \mathcal M$ 
such that $f|_{U_{\rho} } = a_{Q}(U_{\rho} )$.
\end{definition}

The global elements of the locally linear quantum real numbers are given by the functions $a_{Q}$ generated by 
e.s-a. operators $\hat A \in \mathcal M$. It suffices to define algebraic relations between elements of $\AAA$ globally, because $\EsubS(\mathcal M)$ is locally connected and so we can treat functions which are defined on disjoint connected components as if they were globally defined.

The first result follows from the fact that the locally constant functions with values in the rationals form a sub-sheaf $\QQ(\EsubS(\mathcal M))$ of $\AAA(\EsubS(\mathcal M))$ which is dense in $\RsubD(\EsubS(\mathcal M))$ in the metric topology $T$.
\begin{theorem}\cite{adelman2} \label{TH4}
$\AAA(\EsubS(\mathcal M))$  is dense in $\RsubD(\EsubS(\mathcal M))$ in the metric
topology $T$.
\end{theorem}
Therefore for a quantum real number given by a section $f$ over the open set $U$ and an integer $n$ there exists an open cover $U_{j}$ of $U$ such that for each $j$ there is a locally linear function $a^{j}_{Q} : U_{j} \to \RR$ such that $|f|_{U_{j}} - a^{j}_{Q}(U_{j})| < 1/n$.

The sheaf $\AAA(\EsubS(\mathcal M))$ inherits the orders $\leq$ and $<$ from
$\QQ(\EsubS(\mathcal M))$.
On the other hand $\AAA(\EsubS(\mathcal M))$ can be ordered as a consequence
of the orders on the e.s-a. operators in $\mathcal M$:
$\hat A$ is \emph{strictly positive}, $\hat A>0$, if $\langle \hat Au,u\rangle  > 0$ for
$u\neq0$, $u\in \mathcal M$, iff $Tr \hat \rho\hat A > 0, \; \forall \hat \rho \in \EsubS(\mathcal M)$. $\hat A$ is \emph{non-negative}, $\hat A\geq{0}$, if $\langle \hat Au,u\rangle  \geq{0}$ for all
$u\in \mathcal M$  iff $Tr \hat \rho\hat A \geq 0, \;  \forall \hat \rho \in \EsubS(\mathcal M)$.

Algebraically, $\AAA(\EsubS)$ is a \emph{residue field},\cite{johnstone}, in the sense that for all $a_{Q}$ 
in $\AAA(\EsubS)$, if $a_{Q}$ is \emph{not invertible} then $a_{Q} =0$, that is, $\lnot (\exists b_{Q} \in \AAA(\EsubS), \; a_{Q}\cdot b_{Q} = 1) \Rightarrow (a_{Q}=0)$. Because $\lnot (\exists b_{Q} \in \AAA(\EsubS), \; a_{Q}\cdot b_{Q} = 1)$ means, for every $W \in \mathcal{O}(\EsubS)$ and for every locally linear qr-number $b_{Q} $, that $a_{Q}(W)\cdot b_{Q}(W) = 0$ that is  $a_{Q}(W) = 0$ for every open set $W$.

Moreover $\AAA(\EsubS)$ is an \emph{apartness field}, $\forall a_{Q}, \: (a_{Q} > 0 \lor a_{Q} < 0) \Leftrightarrow (\exists b_{Q} \in \AAA(\EsubS), \; a_{Q}\cdot b_{Q} = 1)$, i.e., if $a_{Q}$ is apart from $0$ then it is invertible\cite{stout}. Because if $a_{Q}$ is apart from $0$ then we can assume without loss of generality that $a_{Q} > 0$ which means that there is a rational number $q > 0$ with $q \in {a_{Q}}_{-}$. Therefore $a_{Q}(W) > q$ holds for every $W \in \mathcal{O}(\EsubS)$ whence $b_{Q}(W) = \frac {1}{ a_{Q}(W)}$ is an inverse of $a_{Q}(W)$ for every open set $W$.

\begin{lemma}\label{LM26}
The orders $\geq$ and $>$ on $\AAA(\EsubS(\mathcal M))$ inherited
from $\QQ(\EsubS(\mathcal M))$ are equivalent to those obtained
from those on the essentially self-adjoint operators.
\end{lemma}

\begin{proof}
$\hat A$ is a non-negative essentially self-adjoint operator iff
$Tr \hat \rho\hat A \geq 0 $ for all $\hat \rho$ in $\EsubS(\mathcal M)$, 
i.e. $\hat A \geq 0$ globally. When,
as a Dedekind real number, $a= a_Q  \geq 0$ then
${a}^{+}\subset{0}^{+}$ and ${0}_{-}\subset{a}_{-}$. Globally,
$0^{+}=\{q\in\QQ(\EsubS(\mathcal M)) \mid q>0\}$ and $0^{-}=\{q\in\QQ(\EsubS(\mathcal M)) \mid q<0\}$
so that, if $a= a_Q \geq 0$ then $\hat A \geq 0$ and if $\hat A \geq 0$ then
$a= a_Q \geq 0$.

The positivity order for 
$\hat A$ is equivalent to $\hat A$ being bounded away from zero, i.e.
there exists a rational number $q>0$ such that $\langle u,\hat Au \rangle > q$
for all $u\neq 0 $. This gives the equivalence of the operator
$>$ with the $>$ for Dedekind reals,
because for the latter $a>0$ means globally that
$\exists  q \in  \QQ(\EsubS(\mathcal M))$ with $q\in 0^{+}$ and
$q \in  a_{-}$.
\end{proof}
For example, let $\mathcal{P}(\mathcal{M})$ denote the set of all projection operators satisfying $\hat P = \hat P^{\ast} = \hat P^{2}$ then $\mathcal{P}(\mathcal{M})$ is a \emph{poset} (partially ordered set) when
$\hat P^{(1)} \leq \hat P^{(2)}$ is defined by $\hat P^{(1)}\hat P^{(2)} = \hat P^{(1)}$. If $\pi^{(j)}$ is the globally defined qr-number value of the quality $\hat P^{(j)}$ then $\pi^{(1)}(W) \leq \pi^{(2)}(W)$ for all open subsets $W$.

If $a_Q$ and $b_Q$ are globally defined through the essentially self-adjoint operators $\hat A$ and $ \hat B$ respectively, then the qr-number distance between them is given by
the metric $|a_Q-b_Q|$. However the distance between the locally linear quantum numbers can also be determined by the locally linear quantum number $|a -b|_Q$ corresponding to the positive operator, the absolute value of  the essentially self-adjoint operator $\hat A - \hat B$, viz., $|\hat A-\hat B| =((\hat  A - \hat B)^{2})^{\frac12}$.
\begin{proposition}\cite{adelman2}\label{PR27}
The two metrics coincide, $|a -b|_Q = |a_Q-b_Q|$, for locally linear qr-numbers $a_Q$, $b_Q$ and $|a-b|_{Q}$.
\end{proposition}
\begin{proof}
It is sufficient to let $\hat B=0$. We will only
consider the global sections.

It is well-known, see for example section VI.2.7 of Kato\cite{kato}, that
$\norm{(u,\hat Au)}\leq(u,|\hat A|u)$ for all
$u$ in the domain $D(A)=D(|A|)$, whence
$\norm{\trace\rho{A}}\leq\trace\rho|A|$ for all $\rho$
in $\EsubS$, i.e. $|a_{Q}| \leq   |a|_{Q}$.

As elements of $\AAA$,
$|a_{Q}|=\max(a_{Q},-a_{Q})$ and
$\norm{ |a|_{Q}} =\norm {a}_{Q} $. The lower cut of
$|a_{Q}|$ is the union of the lower cuts of
$a_{Q}$ and of $-a_{Q}$, that is
$|a_{Q}|_{-}=(a_{Q})_{-}\cup\,(-a_{Q})_{-}$,
which means that
$ {\norm{a}_{Q}}_{-}\subset |a_{Q}|_{-}$. The upper cut
of $|a_{Q}|$ is the intersection of the upper cuts of $a_{Q}$ 
and $-a_{Q}$, that is
$q\in\QQ$ belongs to $|a_{Q}|^{+}$ if $q$ is greater
than or equal to both $a_{Q}$ and $-a_{Q}$.
Thus if $q\in|a_{Q}|^{+}$ then $q\in \norm{a}_{Q}^{+}$, 
therefore $|a_{Q}|^{+}\subset \norm{a}_{Q}^{+}$.

This shows that $\norm{a}_{Q} \leq\norm{a_{Q}}$ and therefore
$\norm{a}_{Q}=\norm{a_{Q}}$.
\end{proof}

Either metric can be used to define Cauchy sequences in $\AAA(\EsubS(\mathcal M))$. In order to
ensure uniqueness of the limits of a Cauchy sequence we need the concept of \emph{apartness}, $><$. 
\begin{definition}
The pair $(a_{Q},b_{Q})$ of locally linear qr-numbers are apart,
$a_{Q}><b_{Q}$, iff $(a_{Q}>b_{Q})\vee(a_{Q}<b_{Q})$.
\end{definition}
It follows that
\begin{corollary}
The pair $(a_{Q},b_{Q})$  of locally linear qr-numbers are apart,
$a_{Q}><b_{Q}$, iff $\norm{ a_{Q} -  b_{Q} } > 0$, iff $\norm{a-b}_{Q} > 0$.
\end{corollary}

The locally linear quantum real numbers $a_Q$ and $b_Q$ are apart on the open set $W \subset \EsubS(\mathcal M)$ iff $(a_{Q}(W)>b_{Q}(W))\vee(a_{Q}(W)<b_{Q}(W))$ iff $\norm{ a_{Q}(W) -  b_{Q}(W) } > 0$ iff $\norm{ a-b}_{Q}(W) > 0$. 

On the other hand if  $\norm{ a_{Q}(W) -  b_{Q}(W) } = 0$ on the open set $W$ then $\norm{ a-b}_{Q}(W) = 0$ which by Proposition 13  means that $\norm{\hat A - \hat B} =\hat O$ the zero operator,  that is $ \hat A = \hat B$.
\begin{lemma}
The pair $(x_{Q}(W_{0}),x_{Q}(W_{1}))$  of locally linear qr-numbers generated by the essentially self-adjoint operator $\hat X$ are apart iff $W_{0} \cap W_{1} = \emptyset$ and there exists a rational number $q$ such that  $x_{Q}(W_{0}) < q \land x_{Q}(W_{1}) > q$ or $x_{Q}(W_{0}) > q \land x_{Q}(W_{1}) < q$.
\end{lemma}
\begin{proof}  By proposition 18,  $ x_{Q}(W_{0})$ and  $ x_{Q}(W_{1})$ are apart iff $|x_{Q}(W_{0}) - x_{Q}(W_{1})| > 0$ which implies that the open sets are disjoint, $W_{0} \cap W_{1} = \emptyset$, otherwise  on $W_{0} \cap W_{1}$ the separation $ |x_{Q}( W_{0} \cap W_{1}) - x_{Q}( W_{0} \cap W_{1})| = 0$. Furthermore the definition of apartness prohibits the sections from agreeing on subsets of $\EsubS$ that have an empty interior because there has to a rational number $q$ such that  $x_{Q}(W_{0}) < q \land x_{Q}(W_{1}) > q$ or $x_{Q}(W_{0}) > q \land x_{Q}(W_{1}) < q$ for either $ x_{Q}(W_{0}) < x_{Q}(W_{1})$ or $ x_{Q}(W_{0}) > x_{Q}(W_{1})$ to hold.
\end{proof}

\subsubsection{Qr-numbers generated by a single operator}
The construction of the sheaf of Dedekind reals generated by a continuous function is discussed in the appendix $\S B.6$. For the qr-number $a_{Q}$ defined by the e.s-a operator $\hat A \in \mathcal{M}$, we can isolate the sub-sheaf $\AAA^{a_{Q}}$ generated by $\hat A$ from the sheaf of locally linear qr-numbers. Continuous functions of the sections $a_{Q}(W), \;\forall  W \in \mathcal{O}(\EsubS(\mathcal M))$, define the sub-sheaf  $\RsubD^{a_{Q}}(\EsubS(\mathcal M))$ of qr-numbers generated by $a_{Q}$. It forms a continuum of real numbers because the following conditions are satisfied by $\RsubD^{a_{Q}}(\EsubS(\mathcal M))$: (1)  It contains both the integers $\ZZ(\EsubS(\mathcal M))$ and the rational numbers $\QQ(\EsubS(\mathcal M))$, (2) It has partial orders $<$ and  $\leq$ compatible with those on the rationals $\QQ(\EsubS(\mathcal M))$,(3) It is a residue field; if $b = F(a_{Q}(W)) \in \RsubD^{a_{Q}}(\EsubS(\mathcal M))$ is not invertible then it is zero and (4) With the metric $|\cdot|$, it is a complete metric space in which the rationals $\QQ$ are dense. It is an apartness field, i.e., $\forall b\in \RsubD^{a_{Q}}(\EsubS(\mathcal M)), \;  |b| > 0 $ iff $b$ is invertible.

\begin{proposition}\label{PR29} If the closure of the numerical range of $\hat A$ is $\RR$ then $\RsubD^{a_{Q}}(\EsubS(\mathcal M))$ is a continuum of real numbers.
\end{proposition}
\begin{proof}
 1. Given $\hat A$, for each non-empty open subset $W \in \mathcal{O}(\EsubS(\mathcal M))$ there is a locally linear qr-number $a_{Q}(W)$ and for each continuous function $F : \RR \to \RR$ a qr-number $F(a_{Q}(W))$ defined to extent $W$.  In particular 
$\RsubD^{a_{Q}}(\EsubS(U))$ contains the integers $\ZZ(\EsubS(U))$, the rational numbers $\QQ(\EsubS(U))$, and all the standard real numbers, defined as locally constant functions from $U \to \ZZ,\:\text{or}\; \QQ, \;\text{or}\; \RR$. The qr-numbers  $\RsubD^{a_{Q}}(\EsubS(U))$ can be extended by zero to be globally defined.

2. The orders on $\RsubD^{a_{Q}}(\EsubS(\mathcal M))$ are just the restriction of those on $\RsubD(\EsubS(\mathcal M))$.

3. Since the products and sums of qr-numbers are defined pointwise the proofs that $\RsubD^{a_{Q}}(\EsubS(\mathcal M))$ is a residue field are the same as for $\RsubD(\EsubS(\mathcal M))$ \cite{stout}.
Algebraically, $\RsubD^{a_{Q}}(\EsubS)$ is a \emph{residue field},\cite{johnstone}, in the sense that for all $a_{Q}$ in $\AAA^{a_{Q}}(\EsubS)$, if $a_{Q}$ is \emph{not invertible} then $a_{Q} =0$, that is, $\lnot (\exists b_{Q} \in \RsubD^{a_{Q}}(\EsubS), \; a_{Q}\cdot b_{Q} = 1) \Rightarrow (a_{Q}=0)$. Because $\lnot (\exists b_{Q} \in \RsubD^{a_{Q}}(\EsubS), \; a_{Q}\cdot b_{Q} = 1)$ means  that $a_{Q}(W)\cdot b_{Q}(W) = 0$ for every $W \in \mathcal{O}(\EsubS)$ and for every qr-number $b_{Q} $, that is,  $a_{Q}(W) = 0$ for every open set $W$. If $F(a_{Q})$ is not invertible, for a continuous function $F : \RR \to \RR$ then a similar argument reveals that  $F(a_{Q}(W))\cdot b_{Q}(W) = 0$ for every $W \in \mathcal{O}(\EsubS)$ and for every qr-number $b_{Q} $ so that $F(a_{Q}(W) = 0$ for every open set $W$.

4. Restrict the distance function defined on $\RsubD(\EsubS(\mathcal M))$ to $\RsubD^{a_{Q}}(\EsubS(\mathcal M))$. 
Since the closure of $\hat A$'s numerical range, $\Theta(\hat A)$, contains all the rational numbers $\QQ$ then for any ordered pair $( r_{1}, r_{2} )$ of rationals there are infinitely many qr-numbers $a_{Q}(W)$ such that $r_{1} < a_{Q}(W) < r_{2}$. They are obtained by constructing open subsets $W \subset \EsubS(\mathcal M)$ by setting $W = \{ \rho : s_{1} < Tr \rho \hat A  < s_{2} \}$ for any pair of rationals $(s_{1}, s_{2})$ such that   $r_{1} < s_{1} < s_{2}  < r_{2}$. When the interval topology $T$ is restricted to $\RsubD^{a_{Q}}(\EsubS(\mathcal M))$ the sheaf of rational numbers  $\QQ(\EsubS(\mathcal M)$ is dense in $\RsubD^{a_{Q}}(\EsubS(\mathcal M))$ because to show that  $a_{Q}(W)$ can be arbitrarily well approximated with respect to the distance function $d$ we need to find an open cover $\{ W_{i}\}$ of $W$ such that for each  $i$ there is a locally constant function $q_{i} : W_{i} \to \QQ$ with $d(a_{Q}(W_{i}, q_{i}(W_{i}))$ arbitrarily small. This is just the basic construction of a Dedekind real. \end{proof}
The following lemma and its corollary prove completeness.
\begin{lemma}\label{LM 30}
A Cauchy sequence of locally linear qr-numbers 
$\{ a_{Q}(V_{k})\} \in \AAA^{a_{Q}}(\EsubS(\mathcal M))$, for a fixed operator $\hat A$, converges to a locally linear qr-number $a_{Q}(W)$ for some open subset $W \in \EsubS(\mathcal M)$.\end{lemma}
\begin{proof}The sum $a_{Q}(W) + a_{Q}(V)$ is defined pointwise, so that the sum is defined on $W \cup V$ and is equal to $2a_{Q}(W \cap V)$ on $W \cap V$, to $a_{Q}(W \setminus (W \cap V))$ on $W \setminus (W \cap V)$ and to $a_{Q}(V \setminus (W \cap V))$ on $V \setminus (W \cap V)$.
Then the distance between $a_{Q}(V)$ and $a_{Q}(W)$ is non-zero only on the the interior of $ V \triangle W$, the symmetric difference between $W$ and $V$.   $ V \triangle W = (V \cup W) \setminus (V \cap W) = (V \cap W^{\prime})\cup (W \cap V^{\prime})$, where $V^{\prime}$ is the set theoretical complement of $V$.  The distance $ | a_{Q}(V) - a_{Q}(W)| = |a_{Q}(\text{int}(V \triangle W))|$ and is zero on every open set in the complement of $ V \triangle W$.

Therefore a sequence of locally linear qr-numbers $\{ a_{Q}(V_{k})\}$ is Cauchy if, given $\epsilon > 0,\; \exists K \in \NN $ such that for all $k, k^{\prime} > K$, 
$|a_{Q}(V_{k}) - a_{Q}(V_{k^{\prime}})| < \epsilon$. This implies that the interior of $V_{k} \triangle V_{k^{\prime}}$ is an open set $\Omega(k,k^{\prime})$ such that $|a_{Q}( \Omega(k,k^{\prime})| < \epsilon $ for all $k, k^{\prime} > K$.  Therefore for each $k > K,\; \exists W \in \mathcal{O}(\EsubS(\mathcal M)) \;  : V_{k} = 
W \cup \ U_{k} $ with $W \cap \ U_{k} = \emptyset$ and $| a_{Q}(\ U_{k})| <\epsilon$. Therefore the sequence $\{ a_{Q}(V_{k})\}$ converges to the locally linear qr-number $a_{Q}(W)$. If the interior of the intersection $\cap_{k > K}V_{k}$ is the open set $W \ne \emptyset$ then the limit of the sequence $\{ a_{Q}(V_{k})\}$ is the locally linear qr-number $a_{Q}(W)$.
\end{proof}
\begin{corollary} For any continuous function $F : \RR \to \RR$  if the sequence $F(a_{Q}(V_{k}))$ is Cauchy then it converges to $F(a_{Q}(W))$ whenever the sequence $\{ a_{Q}(V_{k})\}$ converges to $a_{Q}(W)$.
\end{corollary}
\begin{proof}
For any continuous function $F : \RR \to \RR$  if the sequence $F(a_{Q}(V_{k}))$ is Cauchy then the sequence $a_{Q}(V_{k})$ is also Cauchy because $|a_{Q}(V_{k}) - a_{Q}(V_{k'})| <  2\delta $ for $k,k' > K$ if $|F(a_{Q}(V_{k})) - F(a_{Q}(V_{K+1}))| < \epsilon$ when $|a_{Q}(V_{k}) - a_{Q}(V_{K+1})| <  \delta$. By varying $\epsilon$, the number $\delta= \delta(\epsilon,a_{Q}(V_{K+1}))$ can be controlled so that the sequence $a_{Q}(V_{k})$ converges to $a_{Q}(W)$ for some open set $W$. Therefore the sequence $F(a_{Q}(V_{k}))$ converges to $F(a_{Q}(W))$.   \end{proof}

\subsubsection{Infinitesimal qr-numbers}
We will write $\mathcal M$ in place of  $ dU(\mathcal{E}(\mathcal{G}))$.
The infinitesimal qr-numbers are constructed as a sheaf on $\EsubS(\mathcal M)$ by prolonging by zero a sheaf defined on a subset of $\EsubS(\mathcal M)$ with empty interior. The most important example occurs when the subset is a singleton set containing a state $ \rho_0  \in \EsubS(\mathcal M)$. 

The definition of infinitesimal qr-numbers is approached via the sheaf $\AAA(\EsubS(\mathcal M))$ of locally linear functions on $\EsubS(\mathcal M)$, it is based on the definition of infinitesimal Dedekind real numbers.
\begin{definition}
Let $A$ be a non-empty subset of $\EsubS(\mathcal M)$ with empty interior, then the sheaf $IQ_D(A)$ is composed of germs of locally linear functions from $\EsubS(\mathcal M)$ to $\RR$ restricted to $A$. $IQ_D(A)$ can be uniquely\cite{swan} extended to a sheaf, $IQ^{0}_D(A)$, over $\EsubS(\mathcal M)$ by prolongation by zero. Then  $IQ^{0}_D(A)$ is the sheaf of infinitesimal locally linear qr-numbers on $\EsubS(\mathcal M)$ with support $A$.
\end{definition} 
The set $A$ must not be empty because when $A = \emptyset$ then $IQ^{0}_D(\emptyset) = 0$, the constant sheaf at 0.

In $\Shv (\EsubS(\mathcal M))$, the object $IQ^{0}_D(A)$ has sections $a_{Q}^{0}(U)$ over the open set $U$,  given by a locally linear function $a_Q$ on the set $ (U \cap A)$ and by the constant function $0$ over $U - (U \cap A)$. 
Since all functions in both $IQ^{0}_D(A)$ and $\AAA(\EsubS(\mathcal M))$ are defined on open subsets of $X$, 
the ring of extended locally linear qr-numbers has a sum given by $a_{Q}^{0}+ b_{Q}^{0}$ whose domain is $(U \cap V)$ when $U$ is the domain of $a_{Q}^{0}$ and $V$ is the domain of $b_{Q}^{0}$. Therefore the sum of two locally linear qr-numbers is just their locally linear qr-number sum and  the sum of infinitesimal locally linear qr-numbers is an infinitesimal locally linear qr-number. 

Similarly the product of two functions $a_{Q}^{0}$ and $b_{Q}^{0}$ in the ring of extended locally linear qr-numbers is given by the point-wise product defined on $(U \cap V)$ when $U$ is the domain of $a_{Q}^{0}$ and $V$ is the domain of $b_{Q}^{0}$. Therefore the product of a locally linear qr-numbers and an infinitesimal locally linear qr-numbers is an infinitesimal locally linear qr-numbers. The infinitesimal locally linear qr-numbers form an ideal in the ring of extended locally linear qr-numbers. 

Since the logical extent to which these sums and products exist must be an open set, it is given by the interior of the domain of the sum or product. The logical extent to which $a_{Q}^{0} \cdot b_{Q}^{0} = 0$ is the largest open set on which $a_{Q}^{0}(\rho) b_{Q}^{0}(\rho) = 0$. When $a_{Q}^{0}  \in IQ_D(A)$, $a_{Q}^{0}(\rho)^2 \ne 0$ for all $\rho \in A$. But $A$ has empty interior, so that the logical extent to which  $a_{Q}^{0}(\rho)^2 \ne 0$ is $\emptyset$, therefore the statement $\lnot [a_{Q}^{0}(\rho)^2 \ne 0]$ is true for all $\rho \in \EsubS(\mathcal M)$.   This proves the following result which uses the intuitionistic logic of Dedekind real numbers. 
\begin{theorem}\label{TH5}
Each  infinitesimal locally linear qr-number is an \emph{intuitionistic nilsquare infinitesimal}\cite{jlbell} in the sense that it is not the case that it is not a nilsquare infinitesimal. 
\end{theorem} 
\begin{corollary}\label{CR31}
If $A = \{ \rho_0 \}$ and  $a_{Q}$ is a locally linear function, then  $a_{Q}^{0}(\rho_0) \in IQ^{0}_D(\{  \rho_0\})$ is an  infinitesimal locally linear qr-number which is an intuitionistic nilsquare infinitesimal for the Dedekind real numbers $\RsubD(\EsubS(\mathcal M))$.
\end{corollary}
Since every qr-number is a continuous function of locally linear functions\cite{adelman2} each infinitesimal qr-numbers is given by a function $F$ of an infinitesimal locally linear function, if the Maclaurin series for $F$ has no constant term.
\begin{theorem}\label{TH6}
Let $F$ be such a function from $\AAA(\EsubS(\mathcal M))$ to $\RsubD(\EsubS(\mathcal M))$, that maps sections of  $\AAA(W)$ to sections of $\RsubD(W)$ for some open set $W \subset \EsubS(\mathcal M)$. If, for $A \subset W$, $a_{Q}^{0} \in IQ^{0}_D(A)$ is an infinitesimal locally linear qr-number, i.e., if $A$ has empty interior, then $F(a_{Q}^{0} )$ is also an intuitionistic nilsquare infinitesimal qr-number.  
\end{theorem}
\begin{proof}
Any infinitesimal locally linear qr-number $a_{Q}^{0}$ is an intuitionistic nilsquare infinitesimal,i.e.,  $\lnot [a_{Q}^{0}(\rho)^2 \ne 0]$ is true for all $\rho \in \EsubS(\mathcal M)$. Now by assumptions on $F$, $F(a_{Q}^{0}(\rho))^2= [a_{Q}^{0}(\rho)]^2 G(a_{Q}^{0}(\rho)$ where $G$ is a continuous function. Therefore $\lnot [F(a_{Q}^{0}(\rho))^2 \ne 0]$ is true for all $\rho \in \EsubS(\mathcal M)$.
\end{proof}
The expectation values $Tr \hat \rho \hat A$ are infinitesimal qr-numbers for any state $\hat \rho$ and any self adjoint operator $A$ in the algebra $\mathcal M$. All positive powers of an expectation value evaluated at the state $\hat \rho$ are infinitesimal qr-numbers. If we accept the standard interpretation of expectation values as expectation values of repeated experiments then the
expectation values are given by infinitesimal qr-numbers. The expectation values only determine the infinitesimal structure of the qr-number world. 
  
\subsubsection{The time evolution of infinitesimal qr-numbers}

In the standard Hilbert space model of quantum phenomena if a quality is represented by the operator $\hat A$ at time $t = 0$ then at a later time $t$ it will have evolved to the quality represented by the operator $\hat A(t) = \exp(\frac{\imath t}{\hbar} \hat H  )\hat A\exp(- \frac{\imath t}{\hbar} \hat H )$, where $\hat H$ is the Hamiltonian operator. This implies that the infinitesimal qr-number $a_{Q}^{0}(\rho) = Tr \rho \hat A$ of the quality $\hat A$ at the state $\rho \in  \EsubS(\mathcal{E}(\mathcal{G}))$ evolves to the infinitesimal qr-number $a(t)_{Q}^{0}(\rho) = Tr \rho \hat A(t)$. 

When $\hat H$ belongs  to the O$^{\ast}$- algebra $ dU(\mathcal{E}(\mathcal{G}))$ and if 
$ \dot a_{Q}^{0}(\rho) $ corresponds to the operator $ \frac{\imath}{\hbar} [\hat H, \hat A] $ which also belongs to $dU(\mathcal{E}(\mathcal{G}))$ then we can express this law of motion as a differential equation for the infinitesimal qr-numbers, 
\begin{equation}
\frac{d}{dt}a(t)_{Q}^{0}(\rho) = \dot a(t)_{Q}^{0}(\rho) 
\end{equation}
This description of the evolution of infinitesimal qr-numbers comes from the Heisenberg picture.

The infinitesimal qr-number $a_{Q}^{0}(\rho_{0})$ is defined at the state $\rho_{0}$ and if  $\exp (- \frac{ \imath t}{\hbar} \hat H)$ maps the domain $\mathcal D$ into itself then we can interpret the infinitesimal qr-number $a(t)_{Q}^{0}(\rho_{0})$ as corresponding to the operator $\hat A$ defined at the evolved state $\rho_{t} =  \exp(- \frac{ \imath t}{\hbar} \hat H  )\rho_{0}\exp(\frac{\imath t}{\hbar} \hat H )$. If both $\rho_{t}$ and $ \dot \rho_{t} = \frac{\imath}{\hbar} [\rho_{t}, \hat H] $ belong to $ \EsubS(\mathcal{E}(\mathcal{G}))$, then the quantum state satisfies
\begin{equation}
\frac{d}{dt}\rho_{t} = \dot \rho_{t} 
\end{equation}
Then the solution of the equations of motion for the infinitesimal qr-number $ a(t)_{Q}^{0}(\rho) $ can be expressed as 
\begin{equation}
a(t)_{Q}^{0}(\rho) = a(0)_{Q}^{0}(\dot \rho_{t})
\end{equation}
This description of the evolution of infinitesimal qr-numbers comes from the Schr\"odinger picture. 

The Heisenberg picture describes motion that occurs along the fibre over $\rho$,  the Schr\"odinger picture describes a flow in the underlying state space $\EsubS(\mathcal M)$. The time evolution is vertical along a single fibre in the Heisenberg picture and horizontal across the fibres in the Schr\"odinger picture.

The qr-number $a_{Q}(W)$ is the union of the infinitesimal qr-numbers $a_{Q}^{0}(\rho)$ for $\rho \in W$ so that if we assume we can interchange the processes of taking the limit with that of forming the union we may guess that the time evolution of the qr-number $a_{Q}(W)$ satisfies $\frac{d}{dt}a(t)_{Q}^{0}(W) = \dot a(t)_{Q}^{0}(W)$
where $ \dot a_{Q}^{0}(W) $ is the qr-number value of the operator $ \frac{\imath}{\hbar} [\hat H, \hat A] $ at  $W$. This is not true, in general $\frac{d}{dt}a(t)_{Q}^{0}(W) \neq \dot a(t)_{Q}^{0}(W)$. 

We will investigate the dynamical laws for the qr-number values of a particle's position and momentum in the next sections, where we propose that quantum particles
obey the simplest generalisation of classical dynamics that can be
derived on the assumption that qualities take qr-number values.

\subsection{The covariance of qr-number values of qualities}

 Given a quantum system described by the standard Hilbert space quantum formalism 
 in which there is a quantum state space $\EsubS$ we can construct qr-numbers for the 
 physical quantities of the system. The construction depends upon the following assumptions.
\smallbreak
(0) A system always has a complete quantum state, given by
 a non-empty open subset of standard quantum state space $\EsubS$.
The open subsets of  $\EsubS$ are denoted $ \mathcal O(\EsubS)$.
\smallbreak
(1) At all times the system's  physical quantities have values given by qr-numbers.
If a quantity is represented by a self-adjoint
operator $\hat A$ then when the system is in the complete quantum
state $W \in \mathcal O(\EsubS)$, its qr-number value is $a_Q(W)$, the continuous function from $W$ to $\RR$ whose graph is 
$\{ (\hat \rho ,Tr \hat \rho \hat A) : \hat \rho \in W \}$. Therefore as the
system is assumed to always have a complete quantum state each of its
quantities always has a qr-number value.
\smallbreak

A symmetry of a quantum system is represented as a transformation of the qualities 
of the system under which the laws obeyed by the system are invariant. In the standard 
Hilbert space formulation of quantum mechanics a symmetry is realized by an operator $\mathcal{U}(g)$ 
that acts on the vectors in the Hilbert space of the system; if the system is in a state represented by a vector $\phi$ then, under the symmetry labeled by $g$, $\phi \to \mathcal{U}(g)\phi$. Wigner's theorem states that the probabilities of the theory are left invariant if $\mathcal{U}(g)$ is a unitary, or anti-unitary, operator, i.e., $\mathcal{U}^{\ast}(g) = \mathcal{U}^{-1}(g)$. If the symmetries form a Lie group then a projective unitary representation is required because the state of the system corresponds to a ray in the Hilbert space, $\mathcal{U}(g)\mathcal{U}(g') = \exp(\imath \Xi(g,g')) \mathcal{U}(gg')$. A quality of the quantum system is represented by a linear operator on the Hilbert space $\mathcal{H}$ that carries the representation $\mathcal{U}(g)$ of the symmetry group. The group of transformations acts as a group of automorphisms of the algebra of qualities. 

In this paper each $O^{\ast}$-algebra comes from a unitary representation $\hat U$ of a Lie group $G$ on a Hilbert space $\mathcal{H}$. Its $\ast$-algebra is the enveloping algebra $\mathcal{E}(\mathcal{G})$ of the Lie algebra $\mathcal{G}$ of $G$.

Given a unitary representation $U$ of $G$ on $\mathcal{H}$, there is a $\ast$-representation of $\mathcal{E}(\mathcal{G})$ on $\mathcal{D}^{\infty}(U)$ called the \emph{infinitesimal representation} of the unitary representation $U$ of $G$. Each operator $\imath  dU(x)$ is essentially self-adjoint on $\mathcal{D}^{\infty}(U)$ and represents a quality of the system. 
Subgroups of the symmetry group $G$ are used to label the physical qualities. For example, if the system is a single particle with mass $m > 0$ and spin $s$ then its position vector transforms covariantly under the Euclidean group. That is, to represent the quality of a position the triplet of operators $\hat X_{i}\;,  i= 1,2,3$ must satisfy $\mathcal{U}(R, \vec a)\hat X_{i} \mathcal{U}^{-1}(R, \vec a) =\sum_{j=1}^{3}R_{i j}\hat X_{j} + a_{i}$ for $\vec a \in \RR^{3}$ and $R \in SO(3)$. Here we are taking the transformations as "active" in that they correspond to transformation between equivalent classical reference frames. 

The covariance of a quality induces the covariance of its qr-number values. In the example of the position vector, for $W \in  \mathcal O(\EsubS)$, the qr-number values $\vec x_{Q}(W)$ of the position qualities $\hat X_{i} \; , i = 1,2,3 $ transform under a Euclidean transformation $( R, \vec a)$ to
\begin{equation}
\vec x_{Q}(W)^{\prime} =  R \vec x_{Q}(W) + \vec a
\end{equation}
If a quality took the standard real numbers as values, the components  $x_{i}$ would change in the same way to $ \sum_{j=1}^{3}R_{i j} x_{j} + a_{i}$. Therefore the covariance of qr-number values of a quality are described by the same relations as those of the standard real number values of a classical quality.

\subsection{Galilean relativistic quantum mechanics}

The Hilbert space mathematical formalism of Galilean
relativistic quantum mechanics was given by von Neumann 
\cite{vonneumann}.  In it the physical quantities are 
represented by the self-adjoint operators defined on dense subsets 
of Hilbert space which in the case of bounded operators can be extended to the whole 
Hilbert space. These self-adjoint operators belong to an $O^{*}$
algebra $\mathcal M$, a representation of the algebra of observables on the 
carrier Hilbert space of an irreducible representation of the Galilean group; if the 
symmetric product is used they form a real algebra 
$\mathcal M_{sa}$ \cite{inoue}. For example, for the Schr\"odinger representation of
the canonical commutation relations, as for the irreducible representation 
$\{ m, U,s \}$ of the Galilean group, the Schwartz space $\scrS(\RR^3)$ of infinitely differentiable functions of rapid decrease is common dense domain for the representatives of the physical quantities. 

This gives the prototypical example of O$^{\ast}$-algebras.

We take the particle to be associated with an irreducible projective
unitary representation of the Galilean group $G$.

$G$ is parameterised in the following way  
$ g = (b,\vec a, \vec v, R)$:
$b\in \RR$ for time translations, $\vec a \in \RR^{3}$ for space
translations, $\vec v \in \RR^{3}$ for pure Galilean transformations or 
velocity translations, $R \in SO(3)$ for rotations about a point.
Classically  $g \in G$ can be represented as a subgroup of $GL(5,\RR)$,
 \[
\left(
\begin{array}{ccc}
 R & \vec v  & \vec a  \\
0  & 1  & b  \\
 0 & 0  & 1  
\end{array}
\right)
\]

It acts upon the classical space-time coordinates
sending the column vector $(\vec x, t, 1)$ to $(\vec x', t', 1) = (R\vec x + \vec a +
t\vec v , t + b, 1)$. The group law is 
\begin{equation}
g'g = (b', \vec a', \vec v', R') (b,\vec a, \vec v, R)
= (b' + b, \vec a' + R'\vec a + b\vec v', \vec v' + R'\vec v, R'R)
\end{equation}

The 10 generators of its Lie algebra $\mathcal{G}$ are labeled as follows:
$H$ is the infinitesimal generator of the subgroup of time
translations, $P_1, P_2, P_3 $ are the infinitesimal generators of the
subgroup of spatial translations along the three orthogonal axes of the
standard basis, $K_1, K_2, K_3 $ are the infinitesimal generators of the
subgroup of velocity translations along those axes and
$J_1, J_2, J_3 $ are the infinitesimal generators of the
subgroup of rotations around those axes. The Lie brackets are 
\begin{equation}
[J_i,A_j] = \epsilon_{ijk}A_k,\text{for}  \vec A = \vec P, \vec K, \vec J; \;\smallskip
[H, B_i] = 0,\text{for} \vec B = \vec P, \vec J; \;\smallskip [H, K_i] = - P_i.
\end{equation} and
\begin{equation}
[K_i,K_j] = 0, [P_i,P_j] = 0, [K_i,P_j] = 0.
\end{equation}

The enveloping algebra $\mathcal{E}(\mathcal{G})$ of $\mathcal{G}$ has two 
independent invariants in its centre,
\begin{equation}
\vec P \cdot \vec P = \sum_{i=1}^{3} P_{i}^{2} \; \text{and} \; (\vec K \times \vec P)^{2} = \sum_{i=1}^{3} (\vec K \times \vec P)_{i}^{2}
\end{equation}

The projective unitary representations of $G$ are given by unitary representations of the centrally extended Galilean group $G_{m}$, $m>0$, whose elements 
$(\theta, g); \theta \in \RR, g \in G$ obey the group law 
\begin{equation}
(\theta', g')(\theta, g) = (\theta' + \theta + \Xi(g',g), g'g)
\end{equation}
where
\begin{equation}
\Xi(g',g) =  m( \frac{1}{2}b \vec v' \cdot \vec v' + \vec v' \cdot R'\vec a)
\end{equation}
Extended Galilean groups $G_{m}$ for different $m>0$ are isomorphic under the map 
$(\theta, g) \to (\frac{m'}{m} \theta, g)$

The Lie algebra $\mathcal{G}_{m}$ is generated by $ H, \vec P, \vec K, 
\vec J $ and a central element $E$ whose Lie brackets with all the other
10 elements vanish.  We take $E = \hbar I$, then the Lie brackets of the generators of $\mathcal{G}$
remain unchanged except for $[K_i,P_j] = m \delta_{ij}\hbar I.$ 
The enveloping algebra $\mathcal{E}(\mathcal{G}_{m})$ of $\mathcal{G}_{m}$ has two invariants
as well as $m\hbar I$,
\begin{equation}
 U:= H - \frac{1}{2m} \vec P\cdot \vec P \; \text{and} \; \vec S^{2} =  ( \vec J - \frac{1}{m}\vec K \times \vec P)^{2}
\end{equation}

The irreducible projective unitary representations of the Galilean group
$G$ \cite{levy} are obtained from the central extension of the covering group of $G$ in which the rotation group $SO(3, \RR)$ is replaced by its covering group $SU(2,\CC)$. These representations are labeled by the mass $m$, internal energy $U$ and spin $s$, where $m > 0$  and
$U$ are standard real numbers and $s$ is a natural
number or half a natural number. The representation labeled $(m, U, s)$ acts on the Hilbert
space $\scrH :=
\mathcal{L}^{2}(\mathbb{R}^3) \otimes \mathbb{C}^{2s +1}$. The elements of
$\scrH$ are $(2s + 1)$-component vectors of square integrable functions
$\vec x \in \RR^{3} \mapsto \{ \psi_{i} (\vec x) \}, i = -s,..,s $. The
corresponding space-time functions $\psi_{i} (\vec x, t)$ are defined
using the generator $H = {1\over 2m} \vec P\cdot \vec P + U$ of the time
translations,
$\psi_{i} (\vec x, t) := (exp(-\imath Ht/\hbar) \psi_{i}) (\vec x)$.  Then
\begin{equation}
\mathcal{U}(b,\vec a, \vec v, R)\psi_{i} (\vec x, t) = exp[\imath
\alpha_{m} (b,\vec a, \vec v, R ; \vec x, t)] \Sigma_{j} D_{ij}^{s}(R)
\psi_{j}(R^{-1}( \vec x - \vec v(t - b) - \vec a, t - b)
\end{equation}
where $ \alpha_{m}(b,\vec a, \vec v, R ; \vec x, t) := [-{1\over
2}(m/\hbar)\vec v\cdot \vec v (t - b) + (m/\hbar)\vec v\cdot (\vec x - \vec a )]$ and
$D_{ij}^{s}(R)$ are the matrix elements of the irreducible projective
representation of $SO(3)$ on $\mathbb{C}^{2s +1}$.

The vector position for the particle is $ \vec X =
{1\over m} \vec K$, the  vector orbital angular momentum is $ \vec L = \vec X \times \vec P $. 
The spin vector is $\vec S =
\vec J - \vec L$. The $\vec S \cdot \vec S$ commutes with all the
elements of the Lie algebra  and in any irreducible unitary
representation the representative of $\vec S \cdot \vec S = s(s + 1)$ with $s$ an integer or
half-integer.

In the irreducible projective representation $(m, U, s)$ with $m>0$, the position operators 
$\hat X_j,  j =1,2,3,$ are given by the
multiplicative operators $\hat X_j \psi_{i} (\vec x, t) = x_j \psi_{i} (\vec x, t)$ while  the 
momentum operators  $\hat P_j,  j =1,2,3,$ are given by the differential operators
 $\hat P_j \psi_{i} (\vec x, t) = -\imath\hbar{\partial \over \partial x_j} \psi_{i} (\vec x, t)$.
 
The operators $\hat{X_{1}}, \hat{X_{2}}, \hat{X_{3}} $ are unitarily
equivalent to each  other, therefore it suffices to
examine the quantum real number values of one of them.
For simplicity we will write $\hat X$ for $\hat{X_{1}}$, 
$\hat{Y}$ for $\hat{X_{2}}$ and $\hat{Z}$ for $\hat{X_{3}}$.
  
The qr-number values of the z-coordinate of the particle  are
given by the sections $\{z_{Q}(U)\}$ where $U$ can be any open subset of
state space $\EsubS$. If 
$U_{\alpha} = N(\hat \rho_{\alpha}, \hat{Z}, \epsilon)$, where 
$ Tr \hat \rho_{\alpha} \hat{Z} = \alpha $ and 
$N(\hat \rho_{\alpha}, \hat{Z}, \epsilon)$ = $ \{ \rho :
|{Tr\hat \rho \hat{Z} - Tr\hat \rho_{\alpha} \hat{Z}} |<
\epsilon\}$, then the qr-number  $\{z_{Q}(U_{\alpha})\}$ is
approximately equal to the standard real number $\alpha$ if the standard
number $\epsilon$ is small. On the other hand $U$ need not be connected,
for example if
$ U = U_{\beta}\cup U_{\alpha}$, a union of open sets similar to
$U_{\alpha}$, then $ Z_{Q}(U)$ could be approximately equal to $\alpha$
on $U_{\alpha}$ but approximately equal to $\beta$ on $U_{\beta}$. It is
easy to construct examples of qr-numbers  $ z_{Q}(U)$ that do
not satisfy trichotomy, for example, take $U = U_{\alpha}$, as above,
with $\alpha = 0$ then $ z_{Q}(U) > 0 \lor  z_{Q}(U) = 0 \lor 
z_{Q}(U)< 0 $ does not hold no matter how $U$ is shrunk by making 
$\epsilon$ smaller. This occurs because 
$\{\rho : Tr \hat{\rho} \hat{Z} = 0\}$ has empty interior \cite{adelman2}.
Therefore the order $ < $ on the quantum real numbers values 
$\{ z_{Q}(U) ; U \in  \mathcal O(\EsubS) \}$ is not total.   

\section{Appendix: Some Mathematical terms}
\subsection{Functional Analysis}\label{fanalysis}
\subsubsection{Linear Operators on a Hilbert space $\scrH$.}\label{ops}
A good introduction is Volume I of Methods of Modern Mathematical Physics  by M. Reed and B. Simon \cite{reed}.

Throughout this paper, we assume that each Hilbert space $\scrH$ is separable. 

A \emph{linear operator} on a Hilbert space $\scrH$ is a linear map from its domain $\scrD$, a linear subspace of $\scrH$, into $\scrH$. We assume that $\scrD$ is dense in $\scrH$. We write $\scrD(A)$ for the domain of the operator $\hat A$.

An operator $\hat B$ is \emph{bounded} if for all $u \in \scrD$, there is a constant $C > 0$ such that $\|\hat Bu\| \leq C\|u\|$.  Bounded operators are \emph{continuous}. A bounded operator $\hat B$ can be extended from its dense domain $\scrD$ to a bounded operator defined on all of $\scrH$ with the same bound $C$. An \emph{unbounded linear operator} $\hat A$ cannot be continuous at any $u \in \scrD$. 

The \emph{graph} $\Gamma(\hat A)$ of a linear operator is the subset of $\scrH \times \scrH$ given by $\{ ( u, \hat Au ) ; \; u \in \scrD(\hat A) \}$.
A linear operator $\hat A$ is \emph{closed} iff its graph is a closed subset of $\scrH \times \scrH$, i.e., if $u_{n} \to u$ and $\hat A u_{n} \to v$ then $\hat A u = v$. 

An operator $\hat A$ has an extension $\hat A' $ if $\Gamma(\hat A) \subset \Gamma(\hat A')$.
An operator $\hat A$ is \emph{closable} if it has a closed extension. The smallest closed extension of a closable operator $\hat A$ is called its \emph{closure}, denoted $\overline{\hat  A}$.

Let $\hat A$ be a densely defined linear operator on $\scrH$, its \emph{adjoint} $\hat A^{\ast}$ is a linear operator with domain $\scrD(\hat A^{\ast})$ the set of $u \in \scrH$ for which $\exists \; w \in \scrH$ with
\begin{equation}
\langle \hat A v , u \rangle = \langle v, w \rangle , \; \forall \; v \in \scrD(\hat A).
\end{equation}
If $\scrD(\hat A^{\ast})$ is dense in $\scrH$, then define $\hat A^{\ast\ast} = (\hat A^{\ast})^{\ast}$.

If $\hat A$ has a dense domain then $\hat A^{\ast}$ is closed and $\hat A$ is closable iff $\scrD(\hat A^{\ast})$ is dense in which case $\overline{ \hat A} = \hat A^{\ast\ast}$. Furthermore if $\hat A$ is closable then $(\overline{ \hat A})^{\ast} = \hat A^{\ast}$.

A linear operator $\hat A$ with a dense domain is \emph{symmetric} (Hermitian) iff  $\hat A^{\ast} $ is an extension of $\hat A$.
Then $\langle \hat A v , u \rangle = \langle v, \hat A u \rangle, \; \forall \; u, v \in \scrD(\hat A).$

A symmetric operator $\hat A$ with a dense domain is \emph{self-adjoint} iff  $\hat A^{\ast} = \hat A$, that is, iff $\hat A$ is symmetric and $\scrD(\hat A) = \scrD(\hat A^{\ast})$.
 
A symmetric operator $\hat A$ with a dense domain is \emph{essentially self-adjoint} iff  its closure $\overline{\hat A}$ is self-adjoint, that is iff $\overline{\hat A}^{\ast} = \overline{\hat A}$. If $\hat A$ is essentially self-adjoint it has a unique self-adjoint extension. \smallskip

The numerical range $\Theta (\hat A)$ of the operator $\hat A$ is the set of all complex numbers $\langle \hat A u, u\rangle$ where $u$ ranges over all $u \in \scrD(\hat A)$ with $\| u \| = 1$, see $\S 3.2$ in \cite{kato}.  $\Theta (\hat A)$ is a convex set and if $\hat A$ is densely defined and symmetric then it is a subset of $\RR$.

\subsubsection{Trace class operators}\label{tr}
Let $\hat A \in \mathcal B(\mathcal H)$ be positive, $\hat A \geq 0$, so that it is self-adjoint and let $\{e_{n}\}_{n=1}^{\infty}$ be an orthonormal basis of $\mathcal H$ then the \emph{trace of $\hat A$} is the number $Tr \hat A = \sum_{n=1}^{\infty}\langle e_{n}, \hat A e_{n}\rangle$. The trace of $\hat A$ is independent of the orthonormal basis chosen, that is,  $Tr \hat A = Tr (\hat U^{\ast} \hat A \hat U)$ for any unitary operator $\hat U$. 
\begin{definition}
An operator $\hat A \in \mathcal B(\mathcal H)$  is called trace class iff $Tr \vert \hat A \vert < \infty$ where $\vert\hat A \vert = \sqrt{ (\hat A^{\ast} \hat A)}$.
\end{definition}
An alternative definition uses Hilbert-Schmidt operators; if $\hat B \in \mathcal B(\mathcal H)$ with
$\{e_{n}\}_{n=1}^{\infty}$ an orthonormal basis, then 
\begin{equation}
|| \hat B ||_{2} = ( \Sigma_{n=1}^{\infty} || \hat Be_{n} ||^{2} )^{{1\over 2}} \leq \infty
\end{equation}
is the Hilbert-Schmidt norm of $\hat B$.  $|| \hat B ||_{2}$ is independent of the o.n.basis.
\begin{definition}
$\hat B$ is a Hilbert-Schmidt (H-S) operator iff $|| \hat B ||_{2}^{2}  < \infty $. $\mathcal B_{2}(\mathcal H)$ denotes the set of all H-S operators. 
\end{definition}
\begin{theorem}
$\hat A \in \mathcal B(\mathcal H)$ is a trace class operator if it can be represented as a product of two Hilbert-Schmidt operators. 
\end{theorem}

The set of all trace class operators on $\scrH$ is denoted $ \scrT_{1}(\scrH)$. Each $\hat T \in \scrT_{1}(\scrH)$ has a canonical representation as $\hat T = \sum t_{n} \xi_{n} \otimes \overline{\eta}_{n}$, where $\{ t_{n}\} \subset \CC$,\;$\sum |t_{n}| < \infty$ and $\{\xi_{n} \}_{n\in N'}$ and $\{\eta_{n} \}_{n\in N'}$ are orthonormal sets of vectors in $\scrH$ with 
$N' = \{ n \in \NN;\; t_{n} \neq 0\} $. The trace norm of $\hat T$ is $\nu(\hat T) = \sum |t_{n}|$. The trace class operator $\hat T = \sum t_{n} \xi_{n} \otimes \overline{\eta}_{n}$ sends the vector  $\psi \in \scrH$ to
$\sum  t_{n} \langle\psi, \eta_{n} \rangle \xi_{n}$, therefore we will sometimes write 
$\hat T = \sum t_{n} \langle\; , \eta_{n}\rangle \xi_{n}$. 

When 
$\hat T$ is self-adjoint, $\hat T = \hat T^{\ast}$, then $t_{n} \in \RR$ and $\xi_{n} = \eta_{n}$ for all $n \in N'$, in addition, we will take $\xi_{n} = \eta_{n} = 0$, the zero vector, for all $n \in \NN \setminus N'$ in the canonical representation of a self-adjoint trace class operator. That is, $\hat T = \sum t_{n} \xi_{n} \otimes \overline{\xi}_{n}$ or $\hat T = \sum t_{n} \langle \; , \xi_{n}\rangle  \xi_{n} = \sum t_{n} \hat P_{n}$ where $\hat P_{n}$ is the orthogonal projection operator onto the one-dimensional subspace spanned by the unit vector $\xi_{n}$
 
\subsubsection{$\ast$-algebras} \label{alg}
An  \emph{involution} on a vector space $V$ over $\CC$ is a map of $V$ into itself given by $v \to v^{+}$ for all $v \in V$ that satisfies;\;(Vi) $(\alpha v + \beta u)^{+} = \overline{\alpha} v^{+} + \overline{\beta} u^{+}$ , and (Vii) $(v^{+})^{+} = v$, \; for all $v,u \in V$ and all $\alpha, \beta \in \CC$. 

 An \emph{algebra} is a vector space $\mathcal{A}$ in which the map $\mu : \mathcal{A} \times \mathcal{A} \to \mathcal{A}$ given by by $\mu(a,b) = ab$ and called the product of $a$ and $b$ satisfies; (Mi) $a(bc) = (ab)c$,\; (Mii) $(a + b)c = ac + bc$ and $ a(b+c) = ab + ac$,\; (Miii) $\alpha(ab) =( \alpha a)b = a ( \alpha b) $, for all $a,b,c \in \mathcal{A}$ and all $\alpha \in \CC$. The element $I \in \mathcal{A}$ that satisfies $Ia = aI = a$ for all $a \in \mathcal{A}$ is called the identity or unit element of $\mathcal{A}$.
 
 A \emph{$\ast$-algebra} is an algebra $\mathcal{A}$ with an involution $a \to a^{+}$ that satisfies $(ab)^{+} = b^{+} a^{+}$ for all $a,b \in A$ as well as (Vi) and (Vii)

\subsubsection{Locally convex spaces}\label{lcs}

Let $E$ be a vector space over $\CC$ (or $\RR$) and let $\RR_{0} = [0,\infty)$. A subset $X \subset E$ is called \emph{convex} iff $x,y \in X $ implies $ \lambda x + (1 - \lambda ) y \; \in X $ for all $0\leq \lambda \leq 1$. For any $Y \subset E$, the \emph{convex hull} of $Y$ is $c(Y) = \cap X; \; Y\subset X$ and $X$ is convex.

A \emph{topological vector space}, TVS, is a vector space $E$ with a topology such that both addition and scalar multiplication are continuous, i.e., the maps $\sigma : E \times E \to E \; ; (x,y) \mapsto x+y$ and 
$\mu : \CC \times E \to E\; ; (\lambda, x ) \mapsto \lambda x$ are continuous. E.g. any Banach space is a TVS and there are many TVS which are not Banach spaces.

A \emph{locally  convex space} is a TVS that has a base of convex neighbourhoods of zero.

We will be interested in TVS whose topology is generated by a family of semi-norms.

A \emph{semi-norm} on $E$ is a map $p: E \to \RR_{0}$ that satisfies for $x,y\in E$ and $\alpha \in \CC\;(\text{or} \RR)$:
(i)$ p(x+y) \leq  p(x) + p(y)$ and (ii) $p(\alpha x) = |\alpha|p(x)$. If, in addition, $p(x) = 0$ implies $x=0$, $p$ is a \emph{norm}. 

A non-empty family of semi-norms $\{ p_{\alpha} \}_{\alpha \in J}$ \emph{separates points} if $p_{\alpha}(x) = 0 $ for all $\alpha \in J$ implies $x = 0$. 

We are interested in the topology on $E$ which is the weakest in which all the functions $p_{\alpha} $ are continuous and in which the operation of addition is continuous. This is the topology generated by the family of semi-norms $\{ p_{\alpha} \}_{\alpha \in J}$. It is Hausdorff if the family separates points. We will only use families that separate points. A \emph{neighbourhood base at $0$} for this topology is given by the sets $\{ \mathcal N_{\alpha_{1},.....,\alpha_{n} \; ; \epsilon} | \alpha_{j} \in J, j = 1,...,n< \infty, \epsilon > 0 \}$ where 
\begin{equation}
\mathcal N_{\alpha_{1},.....,\alpha_{n} \; ; \epsilon} = \{ x | p_{\alpha_{j}}(x) < \epsilon, \; j= 1,.....,n \}.
\end{equation}
The sets $\{ \mathcal N_{\alpha_{1},.....,\alpha_{n} \; ; \epsilon}\}$ are convex, therefore $E$ is a locally convex space with the topology generated by the this family of semi-norms.

Two families of semi-norms are \emph{equivalent} if they generate the same topology, which occurs when each map in one family is continuous in the topology generated by the other family and vice versa.

A family of semi-norms $\{ p_{\alpha} \}_{\alpha \in J}$ on a vector space $E$ is \emph{directed} iff $\forall \alpha, \beta \in J, \; \exists \; \gamma \in J$ and a real number $K >0$ such that, for all $x \in E$,
\begin{equation}
p_{\alpha}(x) + p_{\beta}(x) \leq K p_{\gamma}(x).
\end{equation}

If  $\{ p_{\alpha} \}_{\alpha \in J}$ is a directed family of semi-norms then $\{ \mathcal N_{\alpha\; ; \epsilon}| \alpha \in J, \; \epsilon > 0 \}$ is a neighbourhood base at $0$.  We can always construct a directed family from a given a family of semi-norms  $\{ p_{\alpha} \}_{\alpha \in A}$. Let $B$ be the set of all finite subsets of $A$, then if $F \in B$, let $d_{F} = \sum_{\alpha \in F} p_{\alpha}$. The family $\{ d_{F} \}_{F \in B}$ is directed and equivalent to the family $\{ p_{\alpha} \}_{\alpha \in A}$.
\begin{theorem}\cite{reed}
Let $X$ and $Y$ be locally convex spaces with families of semi-norms $\{ p_{\alpha} \}_{\alpha \in A}$ and $\{ d_{\beta} \}_{\beta \in B}$. Then a linear map $L: X  \to Y$ is continuous iff $\forall \beta \in B, \; \exists \alpha_{j} \in A, j = 1,....,n$ and a constant $K > 0$ so
\begin{equation}
d_{\beta} (Lx) \leq K (p_{\alpha_1}(x) + p_{\alpha_2}(x) + ....+ p_{\alpha_n}(x)) 
\end{equation}
If the $\{ p_{\alpha} \}_{\alpha \in A}$ are directed, $L$ is continuous iff $\forall \beta \in B, \; \exists \alpha \in A$ and a constant $K>0$ with
 \begin{equation}
d_{\beta} (Lx) \leq K  p_{\alpha}(x).
\end{equation}
\end{theorem}
If $\tau$ is a topology on $E$, write $E(\tau)$ for the corresponding topological space.  

A locally convex space $E(\tau)$ is \emph{metrizable} if its topology $\tau$ is generated by a metric. Thence $E(\tau)$ is metrizable iff the topology $\tau$ is generated a countable family of semi-norms which separates points iff $0$ has a countable neighbourhood base. If $\{ p_{k} \}_{k \in \NN}$ separates points and generates the topology $\tau$ then the function $d(x,y)$ defined on $E\times E$ by 
\begin{equation}
d(x,y) = \sum_{n=1}^{\infty} 2^{- n} \bigl\lbrack \frac{p_{n}(x-y)} {1 + p_{n}(x - y)} \bigr\rbrack
\end{equation}
is a metric on $E$ and also generates the topology $\tau$ on $E$.

A net $\{x_{\alpha}\}$  in a locally convex space $E(\tau)$ \emph{converges} to $x \in E$ iff $p_{\gamma}(x_{\alpha} - x) \to 0$ for each $\gamma \in A$.

A net $\{x_{\alpha}\}$  in a locally convex space is \emph{Cauchy} iff $\forall \epsilon > 0$ and for each semi-norm  $p_{\gamma} $ there is a $\alpha_{0}$ such that $p_{\gamma} (x_{\beta} - x_{\delta}) < \epsilon$
for $\beta, \delta > \alpha_{0}$. $E(\tau)$ is \emph{complete} if every Cauchy net is convergent. 

A \emph{Frechet} space is a complete metrizable locally convex space.

\subsubsection{Lie groups, Lie algebras and enveloping Lie algebras.}\label{lie}
A Lie group is a $C^{\infty}$ manifold $G$ with a smooth map $\mu: G\times G \to G$ which gives $G$ the structure of a group. 
Let $G$ denote a real \emph{Lie group}. We will only consider matrix groups. The \emph{Lie algebra} of $G$ is denoted $\mathcal{G}$, it is the tangent space to the manifold $G$ at the identity $e \in G$. Multiplication in the Lie algebra $\mathcal{G}$ is given for any elements $x,y \in  \mathcal{G}$ by the Lie bracket, $[x,y] = xy - yx$, where the multiplication on the right is matrix multiplication. If $x,y \in \mathcal{G}$, then $\text{ad}\; x(y) = [x,y]$.

The \emph{universal enveloping algebra} of $\mathcal{G}$ is $\mathcal{E}(\mathcal{G})$, the quotient algebra of the tensor algebra over  $\mathcal{G}_{\CC}$ by the two-sided ideal generated by the elements $x\otimes y - y \otimes x - [x,y]$, where $x,y \in  \mathcal{G}$. When $\mathcal{E}(\mathcal{G})$ has been equipped with the involution $x \to x^{+} = - x$ it becomes a $\ast$- algebra with unit.

Let $\{ x_{1}, x_{2},.... ,x_{d} \}$ be a basis for $\mathcal{G}$. For a multi-index $n = (n_{1}, n_{2},....,n_{d}) \in \NN_{0}^{d}$, let $|n| = n_{1} + n_{2} + ..... +n_{d}$ and $ x^{n} = x_{1}^{n_{1}} x_{2}^{n_{2}}.... x_{d}^{n_{d}}$ with $x_{k}^{0} = I$ the unit element of $\mathcal{E}(\mathcal{G})$. The elements $x^{n}, \; n \in \NN_{0}^{d}$ form a basis for the vector space $\mathcal{E}(\mathcal{G})$ (the Poincar\'e-Birkhoff-Witt theorem). For $m \in \NN_{0}$, $\mathcal{E}_{m}(\mathcal{G})$ denotes the linear space of elements $x^{n}$ with $ n \in  \NN_{0}^{d}, \; |n| \leq m$. 

An element $a \in \mathcal{E}(\mathcal{G})$ of the form $a = \sum_{k=0}^{m}\sum_{|n| =k} \alpha_{n} x^{n}$ with $n \in \NN_{0}^{d} $, the integer $m \geq 0$ and $\alpha_{n} \in \CC$, is \emph{elliptic} if $m\neq 0$ and $\sum_{|n| = m} \alpha_{n} y^{n} \neq 0$ for all non-zero vectors $y \in  \NN_{0}^{d} $. The \emph{Nelson Laplacian} in the basis  $\{ x_{1}, x_{2},.... ,x_{d} \}$ is $\Delta = x_{1}^{2} + x_{2}^{2} + .... + x_{d}^{2}$, $\Delta \in \mathcal{E}_{2}(\mathcal{G})$ is an elliptic  element of $\mathcal{E}(\mathcal{G})$, as is $(1 -\Delta)^{n}$ for any integer $n\in \NN$(Schm\"udgen \S 10.2.\cite{inoue}).

\subsubsection{Unitary representations}\label{urs}
A \emph{unitary representation} $\hat U$ of $G$ on the Hilbert space $\mathcal{H}$ is a homomorphism 
$g \to \hat U(g)$ of $G$ into the group of unitary operators on $\mathcal{H}$ with $\hat U(e) = \hat I$ such that the map $g \to \hat U(g)\phi$ of $G$ into $\mathcal{H}$ is continuous for each $\phi \in \mathcal{H}$.

Given a unitary representation $U$ of $G$ on $\mathcal H$, a vector $\phi \in \mathcal{H}$ is a 
$C^{\infty}$-vector for $U$ if the map $g \to U(g)\phi$ of the $C^{\infty} $ manifold $G$ into $\mathcal{H}$ is a $C^{\infty}$-map. The set of $C^{\infty}$-vectors for $U$ is $\mathcal{D}^{\infty}(U)$. It is a dense linear subspace of $\mathcal H$ which is invariant under $\hat U(g), \; g \in G$.

The Garding domain $\mathcal{D}_{G}(U)$ of  $\mathcal H$ for $U$ is the linear span of the vectors  $U_{f}\phi$ where $f \in C_{0}^{\infty}(G)$ and $U_{f}\phi = \int_{G}f(g)U(g)\phi d\mu(g)$ where $\mu$ is a left Haar measure on $G$. $\mathcal{D}_{G}(U)$ is dense in $\mathcal H$, $U(g)U_{f}\phi = U_{f(g^{-1} \cdot)}\phi$, $\mathcal{D}_{G}(U)$  is invariant under $U(g);\; g \in G$ and $\mathcal{D}_{G}(U) \subseteq \mathcal{D}^{\infty}(U)$.

When $\mathcal{D}$ is a dense subspace of a Hilbert space $\mathcal{H}$, a \emph{$\ast$-representation} of the Lie algebra $\mathcal{G}$ on is a homomorphism $\pi$ of $\mathcal{G}$ into the linear operators from $\mathcal{D}$ to itself, $\pi(\alpha x + \beta y) = \alpha\pi( x) + \beta \pi( y)$ and $\pi([x,y]) = \pi(x)\pi(y) - \pi(y)\pi(x) $ for $x,\; y \in \mathcal{G}$ and $\alpha,\; \beta \in \RR$. The operator $\pi(x)$ is skew-symmetric  for each $x \in \mathcal{G}$.

$\forall x \in \mathcal{G}$ define the operator $dU(x)$ with domain $\mathcal{D}^{\infty}(U)$ by
\begin{equation}
dU(x)\phi = \frac{d}{dt}U(\exp tx)\phi|_{t=0} = \lim_{t\to0} t^{-1}(U(\exp tx ) - I)\phi, \; \phi \in \mathcal{D}^{\infty}(U)
\end{equation}
$dU(x)$ gives a $\ast$-representation of $\mathcal{G}$ on $\mathcal{D}^{\infty}(U)$ which has a unique extension to a $\ast$-representation of $\mathcal{E}(\mathcal{G})$ on $\mathcal{D}^{\infty}(U)$ called the \emph{infinitesimal representation} of the unitary representation $U$ of $G$. A representation $\pi$ of $\mathcal{E}(\mathcal{G})$ is \emph{G-integrable} if there exists a unitary representation $U$ of $G$ on $\mathcal{H}(\pi)$ such that $\pi = dU$, i.e., $\forall x \in \mathcal{E}(\mathcal{G})\; , \pi(x) \subseteq dU(x)$ and $\mathcal{D}(\pi) = \mathcal{D}^{\infty}(U)$.

If $x \in \mathcal{G}$ then $\partial U(x)$ is the infinitesimal generator of the strongly continuous one-parameter unitary group $t\in \RR \mapsto U(\exp tx )$ on $\mathcal H$. Then Stone's Theorem \cite{reed} states that  $\imath \partial U(x)$ is self-adjoint on the domain consisting of all $\phi \in \mathcal{H}$ for which the strong operator limit $\lim_{t\to0} t^{-1}(U(\exp tx ) - I)\phi$ exists. Clearly  $ \partial U(x)|_{\mathcal{D}^{\infty}(U)} = dU(x)$.
If $\{x_{1},....,x_{d}\}$ is a basis of $ \mathcal{G}$ then $\mathcal{D}^{\infty}(U) = \bigcap_{k=1}^{d} \mathcal{D}^{\infty}\bigl(\partial U(x_{k})\bigr )$.

The operators $dU(x)$, for $x \in \mathcal{E}(\mathcal{G})$, determine the graph topology $\tau_{dU}$ of the infinitesimal representation of $\mathcal{E}(\mathcal{G})$

Note that $\mathcal{D}^{\infty}(U) $ is a Frechet space with the graph topology $\tau_{dU}$.
\begin{theorem}\cite{inoue} Schm\"udgen, Corollary 10.2.4.
Let a be an elliptic element of $ \mathcal{E}(\mathcal{G})$, then $\mathcal{D}^{\infty}(U) = \mathcal{D}^{\infty}(\overline{dU(a)})$ and the graph topology $\tau_{dU}$ on $\mathcal{D}^{\infty}(U) $ is generated by the directed family of semi-norms $\| \cdot \|_{dU(a)^{n}}, \; n \in \NN_{0}$.
\end{theorem}
\begin{theorem} \cite{inoue} Schm\"udgen, Corollary 10.2.12.
Let $\{ x_{1}, x_{2},.... ,x_{d} \}$ be a basis for $\mathcal{G}$ then $\mathcal{D}^{\infty}(U) = \bigcap_{k=1}^{d} \mathcal{D}^{\infty}(\overline{dU(x_{k})})$.
\end{theorem}
\subsubsection{Analytic and semi-analytic vectors}\label{avs}

Let $E$ be a vector space with a semi-norm $\|\cdot\|$ and $\hat A$ be a linear operator from $E$ into itself, $\hat A \in L(E)$. Then $\phi \in E$ is an \emph{analytic vector for $\hat A$} if $\phi \in \mathcal{D}(\hat A^{n})$ for all $n \in \NN_{0}$ and if $\exists M = M(\phi) \in \RR$ such that $\|\hat A^{n}\phi\| \leq M^{n} n!$ for all $n\in\NN_{0}$ and $\phi \in E$ is an \emph{semi-analytic vector for $\hat A$} if $\phi \in \mathcal{D}(\hat A^{n})$ for all $n \in \NN_{0}$ and if $\exists M = M(\phi) \in \RR$ such that $\|\hat A^{n}\phi\| \leq M^{n} (2n)!$ for all $n\in\NN_{0}$.

$\mathcal{D}^{\omega}(\hat A)$ is the set of all analytic vectors for $\hat A$ and $\mathcal{D}^{s\omega}(\hat A)$ is the set of all semi-analytic vectors for $\hat A$, they are linear subspaces of $\mathcal{D}^{\infty}(\hat A) = \bigcap_{n \in \NN} \mathcal{D}(\hat A^{n})$, with $\mathcal{D}^{\omega}(\hat A) \subseteq \mathcal{D}^{s\omega}(\hat A)$.

If $t>0$ and $\phi \in \mathcal{D}^{\infty}(\hat A)$ define
\begin{equation}
\mathit{e}_{t}^{\hat A}(\phi) = \sum_{n=0}^{\infty}\frac{t^{n}}{n!} \|\hat A^{n}\phi\|
\end{equation}
and 
\begin{equation}
\mathit{s}_{t}^{\hat A}(\phi) = \sum_{n=0}^{\infty}\frac{t^{2n}}{(2n)!} \|\hat A^{n}\phi\|
\end{equation}

Then define $\mathcal{D}_{t}^{\omega}(\hat A) = \{ \phi \in \mathcal{D}^{\infty}(\hat A)\; : \mathit{e}_{t}^{\hat A}(\phi) < \infty \}$ and $\mathcal{D}_{t}^{s\omega}(\hat A) = \{ \phi \in \mathcal{D}^{\infty}(\hat A)\; : \mathit{s}_{t}^{\hat A}(\phi) < \infty \}$, they are linear subspaces of $\mathcal{D}^{\infty}(\hat A)$ equipped with semi-norms $\mathit{e}_{t}^{\hat A}(\cdot)$ and $\mathit{s}_{t}^{\hat A}(\cdot)$, respectively. $\phi \in \mathcal{D}^{\infty}(\hat A)$ is an analytic [semi-analytic] vector for $\hat A$ iff $\exists t > 0$ such the power series for $\mathit{e}_{t}^{\hat A}(\phi)$ [$ \mathit{s}_{t}^{\hat A}(\phi) $] converges.

We now consider families of operators. Let $\mathit{X}$ be a set of linear maps of $E$ into itself. Then $\phi \in E$ is an \emph{analytic vector} for $\mathit{X}$ if $\exists M = M(\phi) \in \RR$ such that
 $\|\hat X_{1}....,\hat X_{N}\phi\| \leq M^{n} n!$ for arbitrary elements $\hat X_{1}, ...., \hat X_{n} \in \mathit{X}$ and for all $n\in\NN_{0}$. $\mathcal{D}^{\omega}(\mathit{X}) \subset E$ is the set of analytic vectors for $ \mathit{X}$. For $n \in \NN$ let
 \begin{equation}
\nu_{n}^{\mathit{X}}(\phi) = sup \{ \|\hat X_{1}....,\hat X_{N}\phi\|\; :\hat X_{1}, ...., \hat X_{n} \in \mathit{X} \}, \;\;\phi \in E  
\end{equation}
and $\nu_{0}^{\mathit{X}}(\phi) = \| \phi\|$. 

If $t>0$ and  $\phi \in E $, let 
\begin{equation}
\nu_{n}^{\mathit{X}}(\phi) = \sum_{n=0}^{\infty}\frac{t^{n}}{n!}\nu_{n}^{\mathit{X}}(\phi) 
\end{equation}
then define $\mathcal{D}_{t}^{\omega}(\mathit{X}) = \{ \phi \in E\; : \mathit{e}_{t}^{\mathit{X}}(\phi) < \infty \}$. It is a linear subspace of $E$ with the semi-norm $ \mathit{e}_{t}^{\mathit{X}}(\cdot)$.  

$\phi \in E $ is an analytic vector for the family $\mathit{X}$ iff $\exists M = M(\phi) \in \RR$ such that $\nu_{n}^{\mathit{X}}(\phi) \leq M^{n} n!$ for all $n \in \NN_{0}$ or equivalently if $\exists t>0$ such that the series for $\nu_{n}^{\mathit{X}}(\phi) $ converges. $\mathcal{D}^{\omega}(\mathit{X}) = \bigcup _{t>0} \mathcal{D}_{t}^{\omega}(\mathit{X})$.

Given $\hat A \in L(E)$ and a family $\mathit{X}\subseteq L(E)$, we say that $\hat A$ \emph{analytically dominates}
$\mathit{X}$ if $\mathcal{D}^{\omega}(\hat A) \subseteq \mathcal{D}^{\omega}(\mathit{X})$, that is, if every analytic vector for the operator $\hat A$ is an analytic vector for the family $\mathit{X}$.

\subsubsection{$\ast$-representations of enveloping algebras}\label{reas}

Let $\pi$ be a representation of the enveloping algebra $\mathcal{E}(\mathcal{G})$ and $\{x_{1},....,x_{d} \}$ be a basis of its Lie algebra $\mathcal{G}$. Any $\phi \in \mathcal{D}(\pi)$ is an analytic vector for $\pi$ if it an analytic vector for the family of operators $\mathit{X} = \{ \pi(x_{1}),....,pi(x_{d})\}$ in $L(\mathcal{D}(\pi))$ relative the Hilbert space norm of $\mathcal{H}(\pi)$.

The set of analytic vectors for $\pi$ is $\mathcal{D}^{\omega}(\pi)$ and $\phi \in \mathcal{D}^{\omega}(\pi)$ iff $\exists M$ such that $\| \pi(x_{k_{1}})....\pi(x_{k_{n}})\;\phi \| \leq M^{n} n! $ for arbitrary indices $k_{j}$ chosen from $\{1,2,....,d\}$ and for all $n \in \NN$.

If $\Delta = \sum_{j=1}^{d} x_{j}^{2}$ is the Nelson Laplacian relative to the given basis of $\mathcal{G}$,
then every analytic vector for the operator $\pi(I - \Delta)$ is an analytic vector for $\pi$.

In particular, for each $x\in \mathcal{E}_{2}(\mathcal{G})$ there exists a number $\lambda_{x} > 0$, independent of $\pi$ such that $\|\pi(x)\;\phi\| \leq \lambda_{x} \| \pi(I - \Delta)\;\phi\|$ for all $\phi \in \mathcal{D}(\pi)$.

And there exists a positive number $\alpha$ such that for elements $x_{j} \in \{x_{1},....,x_{d} \}$ in the basis of $\mathcal{G}$ and any $\phi \in  \mathcal{D}(\pi)$,
\begin{equation}
\|\pi(x_{k})\;\phi\| \leq \alpha \|\pi(I - \Delta)\;\phi\| \;\; \text{and}\;\; \|\pi(x_{k})\pi(x_{l})\; \phi\| \leq \alpha^{2} \|\pi(I - \Delta)\;\phi\|
\end{equation}
and 
\begin{equation}
\|ad\; \pi(x_{k_{1}})......ad\;\pi(x_{k_{n}})\;(\pi(I-\Delta))\;\phi\| \leq \alpha^{n} \|\pi(I - \Delta)\;\phi\| 
\end{equation}

\subsection{Topos theory}\label{top}

A spatial topos, $Shv(X)$, is  the category of sheaves on a topological space $X$: its objects are
sheaves over $X$ and its arrows are sheaf morphisms. That is, an arrow is a continuous function that maps a sheaf $F$ to a sheaf $F'$ in such a way that it sends fibres in $F$ to fibres in $F'$, equivalently, it sends sections of $F$ over $U$ to sections of $F'$ over $U$, where $U$ is an open subset of $X$.

A \textbf{sheaf} is a triplet $\{ F, X, p\}$, where $F$, the sheaf, and $X$, the base, 
are topological spaces and $p$, the projection, is a continuous map from $F$ onto $X$. For each $x \in X$ the inverse image $p^{-1} x$ is called the fibre of $F$ over $x$.
Sheaves on $X$ give a way of describing classes of functions on $X$ through their local properties.  If a property holds on the sheaf $F$ then it holds on all the 
subsheaves $F(U)= F|_{U}$, where $U$ is any open subset of  $X$. Also if a 
property holds on each $F(U_{\alpha})$, where $\{ U_{\alpha} \}$ is an 
open cover of $U$, then it holds on $F(U)$. 

Consider the sheaf $C = C(\EsubS(\mathcal{M}))$ of real-valued continuous functions on $\EsubS(\mathcal{M})$.
A \textbf{section} of the sheaf $C$ over the open subset $U$ of $\EsubS(\mathcal{M})$
is a continuous function $s$ from $U$ to $C$ satisfing the condition
that, for all $\rho$ in $U$, the projection of $s(\rho)$ onto
$\EsubS$ is $\rho$. The sheaf construction allows a section $f$
defined on the open set $U$ to be \textbf{restricted} to sections
$f|_{V}$ on open subsets $V \subset U$ and, conversely, the section $f$ on $U$ can
be recovered by \textbf{collating} the sections $f|_{V'}$ where 
$V'$ belongs to an open cover of $U$.
Existence is local existence: $\exists s $,a section on $W$, with a certain property iff there is an open cover $\{W_i\}$ of $W$ such that for each $i$ the section $s |_{W_i}$ has the property.  

 In any topos there is an object, the subobject classifier $\Omega$, of the truth values in the topos. In  $Shv(X)$ it is the sheaf $\Omega U = \{ W | W \subset U, \; W\in \mathcal O(X)\}$, taking any open set to the set of all its open subsets. The logic of $Shv(X)$ is intuitionistic and only is Boolean when $X$ is the one point set or when the topology on $X$ is $\{ \emptyset, X \}$.

\subsubsection{Real Numbers in a Topos}\label{topnos}
In a topos with a natural numbers object $\NN$, the object of integers $\ZZ$ and the object of rational numbers $\QQ$ can be constructed using the sub-object $\NN^{+}$ of positive integers. $\QQ$ inherits an order relation from $\ZZ$ and satisfies the trichotomy condition that for every pair, $p,q$, of rational numbers we have $p < q \; \lor \; p = q\; \lor \;p > q$. $\QQ$ is a field. In such a topos, an object of Dedekind real numbers can be constructed using cuts in $\QQ$. For details of the construction see\cite{maclane} or \cite{johnstone}. 

A  topos $\Shv(X)$ of sheaves on a topological space $X$ has a natural numbers object $\NN$ so that it has a \emph{Dedekind} real numbers object $\RsubD(X)$ with the following characteristics, proven in  \cite{stout}:

1.  It contains both the integers $\ZZ$ and the rational numbers $\QQ$. Its real numbers exist to variable extents given by open subsets of $X$.

2. It has orders $<$ and  $\leq$ compatible with those on the rationals $\QQ$, but the order $<$ is partial rather than total because the trichotomy condition is not satisfied and  $\leq$ is not equivalent to the disjunction of $<$ and $=$.

3. It is a type of field; closed under the commutative, associative,
distributive binary operations  +  and $\times$, has $0 \ne 1$
and if a number $b$ is not invertible then it is zero, $b = 0$.  

4.It has a distance function $|\cdot|$ which defines a metric with respect to which the real numbers form a complete metric space, with the rational numbers $\QQ$ dense in it. A number $b$ is said to be apart from $0$ iff  $|b| > 0$. The real numbers form an apartness field, i.e., $\forall b, \;  |b| > 0 $ iff $b$ is invertible.

These conditions are sufficient to develop a calculus of functions. The standard real number continuum is an example of Dedekind reals $\RsubD(X)$ in which $X$ is the one point space or if the non-empty set$X$ is equipped with the indiscrete topology $\{ \emptyset, X \}$. 

 Topos theory permits us to vary the mathematical framework for the continuum and hence enlarge the interpretation in much the same way as Riemann enlarged the mathematical framework of metric geometry to include metrics different from the Euclidean in order to gain a richer interpretation.

\subsubsection{ Sheaf real numbers}\label{shfnos}

The sheaf $\RsubD(X)$ of real numbers on the topological space $X$ is a family of copies of the field of 
standard real numbers $\RR_{x}$ parametrized by the points $x \in X$ in a continuous way. For any open set $W \subseteq X$, the disjoint union $\coprod _{x \in W} \RR_{x}$  is a space $\RsubD(W)$ whose topology is such that the projection of $\RsubD(W)$ into $X$, which sends $\RR_{x}$ to the point $x \in W$, is continuous and $\acute{e}$tale in that the topology on  $\RsubD(W)$ is horizontal to match the topology on $X$ while the field structure is vertical along the fibre $\RR_{x}$.  

A section $s$ of $\RsubD(X)$ over $W$ is a function that selects for each $x \in W$ a standard real number $s(x) \in \RR_{x}$ in a continuous way, so that, given an open $V \subset W$, each section over $W$ restricts to a section $s$ over $V$. Conversely the section $s$ over $W$ can be recovered by collating the restrictions of $s$ to each $W_j$ in an open cover of $W = \cup_j W_j$. Sections $s$ over $W$ and $t$ over $V$ collate to give a section $s \cup t$ over $W \cup V$ provided that $s = t$ on the overlap $W\cap V$.

The sheaf  $\RsubD(X)$  can be described wholly in terms of the collection of all sections $s$ defined over open subsets $U$ of $X$, together with the operations of restriction and collation. 

\begin{theorem}\label{TH3}
The object $\RsubD(X)$  of Dedekind real numbers in the topos $\Shv (X)$ of sheaves of sets on the topological space $X$ is isomorphic to the sheaf $\mathcal C(X)$ of continuous real-valued functions on the space $X$ \cite{maclane}. \end{theorem}
\subsubsection{Sections}
A continuous function $s$ on $X$ restricts to sections over open subsets of $X$.
Recall that a bundle over a space $X$ is a continuous map $p: Y \to X$ and a cross-section of the bundle is a continuous map $s : X \to Y$ that satisfies $p\circ s = 1$. The inverse image $p^{-1}(x)$ of any $x \in X$ is called the fibre of $Y$ over $x$. If $U$ is any open subset of $X$ the bundle $p_U : p^{-1} U \to U$ is by restriction a bundle over $U$. A cross-section $s$ of the bundle $p_U$ is a continuous map $s : U \to Y$ such that $p\circ s = i $ the inclusion map $i : U \to X$. If we let $\Gamma_{p}U = \{ s | s: U \to Y\; \text{and}\; p\circ s = i \}$ be the set of all cross-sections over $U$, then $\Gamma_{p}$ is a sheaf, the sheaf of cross-sections of the bundle $p$.

The proof that every sheaf is a sheaf of cross-sections of a bundle depends upon the concept of a germ of a function. Two continuous functions $f$ and $g$ have the same germ at a point $x_0$ if $\exists U$ an open neighbourhood of $x_0$ such that $f(x) = g(x), \; \forall x \in U$. This defines an equivalence relation, the equivalence class of the function $f$ is called, $\text{germ}_{x_0}f$, the germ of $f$ at $x_0$. If, given an open set $U$, $\text{germ}_{x}f = \text{germ}_{x}g$ for all $x \in U$, then, by collation, $f|_U = g|_U$. If $\{ U_{\alpha}\}$ is a set of nested neighbourhoods of $x_{0}$ then for all $f$ the limit of $f|_{U_{\alpha}}$ is  $ f (x_0) \in \RR_{x_0}$ the stalk of the bundle of sections  because if  $\text{germ}_{x_0}f = \text{germ}_{x_0}g$ then $f(x_0) = g(x_0)$.

\subsubsection{Categorical description}\label{cat}
If $X$ is a topological space then $\mathcal O(X)^{op}$ is a category whose objects are open subsets of $X$ and whose arrows are restriction maps $f: V \to U$ if $ U \subset V$ are open subsets of $X$. $\underline{\text{Sets}} $ denotes the category whose objects are small sets and arrows all functions between them. Small sets are subsets of a fixed universe $U$\cite{maclane1}. A functor $F$ from 
$\mathcal O(X)^{op}$ to $ \underline{\text{Sets}} $ is an operation which assigns to each object $U$ of 
$\mathcal O(X)^{op}$ an object $F(U)$ of $ \underline{\text{Sets}} $ and to each arrow $f$ of $\mathcal O(X)^{op}$ an arrow $F(f)$ of $ \underline{\text{Sets}} $ in such a way that $F$ respects the domains and the co-domains as well as the identities and compositions. 

Now we can define a sheaf of sets $F$ on a topological space $X$ is a functor $F : \mathcal O(X)^{op} \to \bf Sets$ such that each open covering $ \cup_{i \in I}U_i = U \in \mathcal O(X)$, gives an equalizer diagram $FU \buildrel e \over \longrightarrow \prod_{i}FU_i {\buildrel p\over \longrightarrow \atop \buildrel q\over \longrightarrow} \prod_{i,j}F(U_i \cap U_j)$, where for $t \in FU, \; e(t) = \{ t|_{U_i} | i \in I \}$ and for a family $t_i \in FU_i$, $p(t_i) = \{ t_i|_{(U_i\cap U_j)}\}$ and $q(t_i) = \{ t_j|_{(U_i\cap U_j)}\}$\cite{maclane}. Then for any point $x \in X$ with $U$ and $V$ two open neighbourhoods of $x$ and two elements $s \in FU$ and $t \in FV$, $s$ and $t$ have the same germ at $x$ if there is an open set $W \subset U\cap V$ with $x \in W$ and $s|_{W} = t|_{W}$. This defines an equivalence relation, the equivalence class of $s$ is called, $\text{germ}_{x}s$, the germ of $s$ at $x$.

 Let $F_x$ be the set of all germs of $F$ at $x \in X$, this is usually called the stalk of $p$ at $x$. $F_x = \lim_{x\in U} FU$ is the colimit of the functor $F$ restricted to open neighbourhoods $U$ of $x$\cite{maclane}.
The disjoint union of the $F_x$ over all points $x \in X$ is $\
u_{F} = \coprod_{x} F_x$. Define $p: \
u_{F} \to X$ as the function that sends each germ, $\text{germ}_{x}s$, to the point $x$.

Then each $s \in FU$ determines a function $\dot s$ by $\dot s: U \to \
u_{P} $ with $\dot s(x) =  \text{germ}_{x}s, \; \forall x \in U$. $\dot s$ is a section of $( \
u_{F}, X, p )$. Topologise the set $\
u_{F}$ by taking the basic open sets to be the image sets $\dot s(U) \subset \
u_{F}$, then each function $\dot s$ and $p$ are continuous. Each $\dot s$ is an open map and an injection, hence $\dot s : U \to \dot s(U)$ is a homeomorphism. The bundle $p : \
u_{F} \to X$ is a local homeomorphism because each point of $\
u_{F}$ has an open neighbourhood that is mapped by $p$ homeomorphically onto an open subset of $X$.

A bundle $p : Y \to X$ is etale when $p$ is a local homeomorphism, i.e., to each $y \in Y, \; \exists V \in \mathcal O(Y)$ with $y \in V$, such that $p(V) \in \mathcal O(X)$ and $p|_V$ is a homeomorphism $V \to p(V)$.  If $p: Y \to X$ is etale then so is the pullback $Y_U  \to U $.

\subsubsection{The Dedekind real numbers $\RsubD(X)$}\label{dedreals}
In $\Shv (X)$, the Dedekind real numbers object $\RsubD(X) \equiv \mathcal C(X)$, so that a section $f|_{U}$ of a continuous function $f$ over an open set $U$ is a variable real number defined to extent $U$. 

This is proved by taking the topos as the "universe of sets", then defining Dedekind real numbers using cuts in the rational numbers as follows: a real number 
$x|_{U} = (x_{-}, x^{+})$ where $x_{-}$ and $x^{+}$ are non-empty subsets of $\QQ(U)$. We write $q,q',..$ for constant functions on $U$ that take the value $q,q',..$. Then the subsets $x_{-}$ and $x^{+}$ of $\QQ(U)$ satisfy

(1) $x_{-} \cap x^{+} = \emptyset$, \;(2)\;$\forall q \in x_{-}, \; \exists q' \in x_{-}, \; q' > q$, \;(3)\; $\forall q \in x^{+}, \; \exists q' \in x^{+}, \; q' < q$, \;(4)\; $\forall q \in x_{-}, \; \forall q' \in x^{+}, \; q <  q' $, \\(5) $\forall n \in \ZZ^{+},\; \exists q \in x_{-}, \; \exists q' \in x^{+}, \; q' - q < 1/n $, where if $n \in \ZZ^{+}$ then $n > 0$.\cite{stout}

If $f|_{U}$ is constant then it is a standard real number defined to extent $U$. $0|_{U}$ is the zero, $1|_{U}$ the unit defined on $U$. $\RsubD(U)$ has integers $\ZZ(U)$, rationals $\QQ(U)$ and Cauchy reals $\RsubC(U)$ as subsheaves of locally constant functions\cite{stout}. A
function is locally constant if it is constant on each connected open subset of its domain.
A continuous function $f$ defined on any subset $V \subset X$ determines a Dedekind real number defined to extent $\text{int}\{V\}$, the interior of $V$, the largest open subset of $V$. 

The propositions in a spatial topos are true to extents given by open subsets of the base space $X$. If $P$ is true to extent $W \in \mathcal O(X)$ and $Q$ is true to extent $V \in \mathcal O(X)$ then the proposition $P \lor Q$ is valid if $W \cup V = X$ and the extent of $ P \Rightarrow Q$ is the largest open set $U$ such that $U \cap W \subset V$ because $ P \Rightarrow Q$ is true iff whenever $P$ is true on an open set $U$, $Q$ is also true on $U$. The negation of a proposition is true on an open set $U$ if the proposition is not true on any open subset of $U$. The language describes local properties: If a proposition is true for a sheaf $F$ then it is true for all subsheaves $F|_{U}$ where $U \in \mathcal O(X)$. If a proposition holds for each $F|_{U_{\alpha}}$, where $\{ U_{\alpha} \}$ is an open cover of $X$, then it is true for $F$.

Real numbers are represented by continuous functions with domains given by open subsets of $X$. Two real numbers are equal, or globally equal, if they are equal to full extent, $X$, they are unequal if they are equal to empty extent, $\emptyset$. Restricting attention to the subsheaf $\RsubD(U)$, if $a$ with domain $W \in \mathcal O(U)$ and $b$ with domain $V \in \mathcal O(U)$ are two real numbers, then $a = b$ to an extent given by an open subset of $W \cap V$.  If $V \subset W$ then $W \cap V = V$ so the numbers $a$ and $b$ can at most be equal to extent $V$.  The sum $a + b$ is defined pointwise, so that $a + b = c$ where $c$ is the continuous function defined on $W \cup V$ which is equal to $a + b$ on $W \cap V$, to $a$ on $W \setminus (W \cap V)$ and to $b$ on $V \setminus (W \cap V)$. The product $a\cdot b$ is defined on $W \cap V$ by $a\cdot b(x) = a(x)b(x); \forall x \in W \cap V$.  If we wish to do analysis with these numbers we have to consider the possibility of forming infinite sums then in order to ensure that the result when it exists is a Dedekind real number we must take its domain to be the interior of the intersection of the domains of the summands.

For any open set $U \subset X$ the usual orders on the
rational numbers $\QQ(U)$ can be extended to $\RsubD(U)$, 
even though trichotomy does not necessarily hold. 

In the following definitions $U$ is a non-empty open subset of $X$.
\begin{definition}
The order relation $<$ on $\RsubD(U)$  is given by:
\begin{displaymath}
x<y\;\text{if and only if}\;\exists\ 
q\in\QQ(U)\;\bigl((q\in{x}^{+})\wedge({q}\in{y}_{-})\bigr)
\end{displaymath}
where $x^{+}$ is the upper cut of $x$ and $y_{-}$ is the
lower cut of $y$. The relation $<$ is the subobject of
$\RsubD(U)\otimes\RsubD(U)$ consisting of
such pairs $(x,y)$.
\end{definition}

\begin{definition}
The order relation $x\leq{y}$ is the
subobject of $\RsubD(U)\otimes\RsubD(U)$
consisting of the pair $(x,y)$ with $y^{+}\subset{x}^{+}$
and $x_{-}\subset{y}_{-}$, where $x^{+}$ is the upper cut
of $x$ and $x_{-}$ is the lower cut of $x$, and similarly
for $y$.
\end{definition}
With the order $\leq$ as defined above,  $x\leq{y}$ is not the same as $(x<y)\vee(x=y)$. However
$(x\leq{y})$ is equivalent to $\neg(y<x)$\cite{stout}, as the second proposition is true if $y<x$ is not true on any open subset.\\ 

Open and closed intervals of $\RsubD(U)$ are given by
\begin{definition}
The open interval $] x, y [$ for
$x<y$ is the subobject of $\RsubD(U)$ consisting
of those $z$ in $\RsubD(U)$ that satisfy $x<z<y$.

The closed interval $[x,y]$ for $x<y$ is the subobject of
$\RsubD(U)$ consisting of those $z$ in
$\RsubD(U)$ that satisfy $x\leq{z}\leq{y}$.
\end{definition}
The open intervals can be used to construct an \emph{interval
topology $T$} on $\RsubD(U)$ analogously to the
interval topology on $\RR$ that is 
generated from the open intervals by finite intersection and
arbitrary union. Then $\QQ(U)$ is dense in $\RsubD(U)$ with respect to the topology $T$\cite{stout}. 

The \emph{maximum function}, $\max$, is defined using the order $\leq$
by the conditions:
(i) $x\leq\max(x,y)$ and $y\leq\max(x,y)$, and
(ii) if $z\leq{x}$ and $z\leq{y}$ then $z\leq\max(x,y)$\cite{stout}. 

A \emph{norm} is defined on $\RsubD(U)$ using the \emph{norm function}
$|\ \ |:\RsubD(U)\to\RsubD(U)$ that
is defined by $|x|=\max(x,-x)$.
The \emph{norm}, denoted  $|\ \ |$, satisfies the usual conditions:
(1) non-negativity, (2) only 0 has norm zero and (3) the 
triangle inequality holds\cite{stout}.

There is an \emph{apartness relation} on $\RsubD(U)$, $a$ is apart from $b$, $ a>< b\; \buildrel\rm def\over= \; (a > b \lor a < b)$. \emph{Apartness} is stronger than 
\emph{not equal to}. Note that $\forall a, b \; (|a - b| > 0) \Leftrightarrow (a > b \lor a < b)$ \cite{stout}.

Topologically,  $(\RsubD(U),T)$ is a \emph{metric space} with metric 
$d(x,y)=|x-y|$. It is both \emph{complete} and \emph{separable}\cite{stout}.

Algebraically, $\RsubD(U)$ is a \emph{residue field},\cite{johnstone}, in the sense that for all $a$ 
in $\RsubD(U)$, if $a$ is \emph{not invertible} then $a =0$, that is, $\lnot (\exists b \in \RsubD(U), \; a\cdot b = 1) \Rightarrow (a=0)$. 
Moreover $\RsubD(U)$ is an \emph{apartness field}, $\forall a, \: (a > 0 \lor a < 0) \Leftrightarrow (\exists b \in \RsubD(U), \; a\cdot b = 1)$, i.e., if $a$ is apart from $0$ then it is invertible\cite{stout}.
 
\subsubsection{Dedekind reals generated by a single function}\label{snglfn}

Let $h$ be a continuous function, $h: X \to \RR$, then define a sub-functor $F_{h} $  from 
$\mathcal O(X)^{op}$ to $ \underline{\text{Sets}} $ by having $F_{h}$ send (i) each open set $ U$ to the continuous function $ h|_{U} : U \to \RR$, which is a one-element subset of $\mathcal C(U)$, and (ii) each restriction map $f : U \to V$ for $V \subset U$ to $F_{h}(f)$ that restricts $ h|_{U}$ to $ h|_{V} ; V \to \RR$. Now by Proposition 1 on page 67 of \cite{maclane} this means that $F_{h} $ is a sub-sheaf of $F$ because for every open $U$ and every element $g \in F(U)$ on every open covering $U = \cup_{j} U_{j}$ we have that  $g\in F_{h}(U)$ iff $g|_{U_{j}} \in F_{h}(U_{j})$ for all $j$. Therefore we can consider the Dedekind real numbers $\RsubD^{h}(X)$ generated by a single continuous function $h: X \to \RR$.

The order relations defined on $\RsubD(X)$, via the usual orders on $\QQ(X)$, can be restricted to $\RsubD^{h}(X)$. If $h$ is not a constant function and $V,W$ are disjoint open subsets of $X$, then $h|_{V} < h|_{W}$ means that for each $v\in V$ and each $w\in W$, $h(v) < h(w)$ and $h|_{V} \leq h|_{W}$ means that for each $v\in V$ and each $w\in W$, $h(v)  \leq h(w)$. The norm is defined on $\RsubD^{h}(X)$ by restriction making it a metric space with the metric $d(h|_{V} ,h|_{W} )=|h|_{V} -h|_{W} |$ for open sets $V$ and $W$. 

Proofs, modeled on those in \cite{stout} that use the properties of Dedekind cuts in $\QQ(X)$, hold when restricted to $\RsubD^{h}(X)$, therefore we have the following
\begin{proposition}\label{PR20}
When the interval topology $T$ is restricted to $\RsubD^{h}(X)$ the sheaf of rational numbers  $\QQ(X)$ is dense in $\RsubD^{h}(X)$.
\end{proposition}
Whether  $\RsubD^{h}(X)$ is complete as a metric space will depend upon the function $h$.
We will return to this question when we discuss  quantum real numbers.
\subsubsection{The extended Dedekind real numbers}\label{exreals}
An alternative way to add and multiply Dedekind real numbers that are defined on different open subsets is to extend each Dedekind real number to be defined on the whole of $X$ using sheaves on $X$ that are prolongations by zero of sheaves defined on non-empty subsets $A$ of $X$. 
\begin{definition}
Let $A$ be a non-empty subset of $X$, if $F$ is any sheaf over $X$ then $F|_{A} = p^{-1}(A)$, is the restriction of  $F$ to $A$ where the projection $p : F \to X$ is a local homeomorphism. Then  $F|_{A} $ is a sheaf over $A$ called the restriction of $F$ to $A$, its projection, topology and algebraic structure are
induced from those of the sheaf $(F, X, p)$.\cite{swan} 
\end{definition}
\begin{definition}
The sheaf $F^{0}(A)$ over $X$ is a prolongation by zero of the sheaf $F|_{A}$ over $A$ if 
$ F^{0}|_{A} = F|_{A}$ and $F^{0}|_{X \setminus A} = 0 $, where $X \setminus A$ is the complement in $X$ of $A$ and $0$ is the constant sheaf at $0$.\cite{swan} 
\end{definition}
We define the \emph{extended} Dedekind real numbers $\RsubD^{0}(X)$.
\begin{definition}
Let $U$ be an open subset of $X$ then the sheaf $\mathcal C(U)$ is composed of germs of continuous functions from $X$ to $\RR$ restricted to $U$. $\mathcal C(U)$ is uniquely\cite{swan} extended to a sheaf, $\mathcal C^{0}(U)$, over $X$ by prolongation by zero. $\mathcal C^{0}(U)$ is the sheaf of extended Dedekind real numbers on $X$ with support $U$.
\end{definition} 

In $\Shv (X)$, the sheaf $\mathcal C^{0}(U)$, for $U$ a non-empty open subset of $X$, has a section 
$f^{0}|_{U}$ over $U$  given by the continuous function $f$ over the set $ U$ and the constant function $0$ over $X \setminus U$.

\subsubsection{Infinitesimal Dedekind real numbers }\label{infdr}
One way of making a continuum flow uses \emph{infinitesimal real numbers} that are not Dedekind real numbers and do not correspond to points but give the difference between neighbouring points.\cite{jlbell} Two points $X$ and $Y$ are neighbours if they are not identical but their coordinated real numbers $x$ and $y$ do not satisfy $ x > y \lor x < y $ to any extent. The difference $(x-y)$ between neighbouring numbers is an \emph{order theoretical infinitesimal number} because there is no open set on which $ x > y \lor x < y $ is true. If the real numbers satisfied trichotomy then the difference between neighbouring real numbers must be zero but our Dedekind real numbers do not satisfy trichotomy.

A promising approach to the construction of infinitesimal Dedekind real numbers comes from the construction of sheaves on $X$ by prolongation of sheaves defined on subsets $A$ whose interiors are empty.  The definition follows.
\begin{definition}
Let $A$ be a subset of $X$ with empty interior, e.g. $A = \{x\}$ contains a single element, then the sheaf $\mathcal C(A)$ is composed of germs of continuous functions from $X$ to $\RR$ restricted to $A$. $\mathcal C(A)$ is uniquely\cite{swan} extended to a sheaf, $\mathcal C^{0}(A)$, over $X$ by prolongation by zero. $\mathcal C^{0}(A)$ is the sheaf of infinitesimal Dedekind real numbers on $X$ with support $A$, when $A$ is a subset of $X$ with empty interior.
\end{definition} 

In $\Shv (X)$, the sheaf $\mathcal C^{0}(A)$, for $A$ a non-empty subset of $X$, has a section 
$f^{0}|_{U}$ over the open set $U$  given by the continuous function $f$ over the set $ (U \cap A)$ and the constant function $0$ over $U - (U \cap A)$. 
When $A$ has an empty interior, there is no open subset $V \subset U$ on which the absolute value of $f^{0}|_{V}$ is positive, therefore $f^{0}|_{U}$ is an\emph{ infinitesimal Dedekind real number} because it is an order theoretical infinitesimal. 

We will restrict our attention to topological spaces $X$ in which each singleton set $\{x\} $ is closed.
If $f$ is a continuous function on such a topological space $X$, then the sections  $f^{0}|_{U}$  and $f^{0}|_{U^{\prime}}$ represent neighbouring points when $U^{\prime} = U - \{x\} = U\cap\{x\}^{\prime} $ for some $x \in U$.  The sheaf of extended Dedekind real numbers $\mathcal C^{0}(X^{\dagger})$ is  described by the collection of all sections both of sheaves $\mathcal C^{0}(\{x\})$ for points $x \in X$ and sheaves $\mathcal C^{0}(U)$ over open subsets $U$ of $X$.

All sections in $\mathcal C^{0}(\{x\})$ and $\mathcal C^{0}(U)$ are constructed from restrictions of continuous functions defined on $X$, therefore the sum $f^{0} + g^{0}$ of the sections $f^{0}$ and $g^{0}$ such that the sum of two Dedekind real numbers is a Dedekind real number, the sum of infinitesimal Dedekind real numbers is an infinitesimal Dedekind real number and the sum of  a Dedekind real number and an infinitesimal Dedekind real number is an extended Dedekind real number. 
The product of (prolonged) Dedekind real numbers is clearly a Dedekind real number.
The product of the sections $f^{0} \in \mathcal C^{0}(U)$ and $g^{0} \in \mathcal C^{0}(\{x\})$ is defined to only be non-zero on $(U \cap \{x\})$. Therefore, if $x \in U$, the product of the Dedekind and infinitesimal Dedekind real numbers is an infinitesimal Dedekind real number. That is, the infinitesimal Dedekind real numbers form an ideal for the ring of extended Dedekind real numbers. 

 Thus each infinitesimal Dedekind real number of the form $g^{0} \in \mathcal C^{0}(\{x\})$ is an \emph{intuitionistic nilsquare infinitesimal}\cite{jlbell} since its product with itself is only non-zero at $\{x\}$ which as a set has empty interior, therefore it is not the case that it is not a nilsquare infinitesimal. 

Another way to see this result is to take a family of open neighbourhoods $\mathcal N(x_0, \epsilon)$ of a point $x_0 \in X$, labelled by a standard real number $\epsilon$, such that in the limit as $\epsilon$ goes to zero we don't get an open set but get the singleton $\{x_0\}$. If we enlarge the field of Dedekind real numbers to include continuous functions restricted to arbitrary intersections of open sets then we get 
infinitesimal Dedekind real numbers.

\subsubsection{Infinitesimal analysis}\label{infan}
The existence of the infinitesimal  Dedekind real numbers permits a development of differential geometry by synthetic reasoning.\cite{kock},\cite{reyes},\cite{jlbell}

There are two classes of functions that we must distinguish: in the first, sections of the sheaf $ \mathcal C(W)$ for the fixed open set $W$ are mapped to sections of the sheaf $ \mathcal C(W)$ for the same open set $W$, in the second sections of the sheaf $ \mathcal C(W)$ may be mapped to sections of the sheaf $ \mathcal C(U)$ for $U \ne W$. The first occurs in physical theories in situations when all the physical quantities take numerical values to a fixed extent and the function converts one quantity into another, in the second the same physical quantity takes numerical values to different extents, the function changes the extents. In quantum theory the first corresponds to the Heisenberg picture, the second to the Schr\"odinger picture.

If $G$ is a continuous function: $\RsubD(W) \to \RsubD(W)$ then the derivative of $G$ at the point whose Dedekind real number coordinate is the continuous real-valued function $f|_{W}$, for $W \in \mathcal O(X)$, is determined by the infinitesimal Dedekind real numbers  $ g|_{x_0}$, where $g$ is any continuous real-valued function on $X$, because the following holds intuitionistically
\begin{equation}
G( f|_{W} +  g|_{x_0}) = G(f|_{W}) +  g|_{x_0}\cdot G'( f|_{ W})
\end{equation}
for any point $x_0 \in X$. If $x_0 \in W$ then the second term is $ g|_{x_0}\cdot G'( f|_{ x_0})$, but if 
$x_0 \notin W$ then the second term vanishes. 

In smooth infinitesimal analysis\cite{jlbell} there are sufficiently many infinitesimals to ensure that the derivative of the function $G$ exists independently of the infinitesimal used in the above equations. This  
is achieved because the equation holds for any real-valued continuous function $g: X \to \RR$. We have the \emph{cancellation law}. Given $x_0 \in X$ and a real-valued continuous function $f : X \to \RR$,
\begin{equation}
g|_{x_0}\cdot G'( f|_{ x_0}) = g|_{x_0}\cdot F'( f|_{ x_0}), \forall g \; \Rightarrow G'( f|_{ x_0}) = F'( f|_{ x_0}) 
\end{equation}

That is, for  Dedekind real numbers $ \RsubD(X) \equiv \mathcal C(X)$, the set of infinitesimals at the point $x_0 \in W \subset X$ are labelled by the set of real-valued continuous function $g: X \to \RR$. If we think of the graph of the continuous function $ y = G(x)$, when $y$ and $x = f|_{W} $ are Dedekind real numbers defined on the open set $W$, i.e., sections of $\mathcal C(W)$, then the bit of the graph above the infinitesimal interval between  $ f|_{W} $ and $f|_{W} +  g|_{x_0}$ is straight and coincides with a line tangent to the graph at the point  $ f|_{W} $. Firstly we move along the stalk over $x_0$ by varying the function $g$, this determines the value $G'( f|_{ x_0})$ by the cancellation law, then we move over the open set $W$ by varying the point $x_0 \in W$ to get the number $G'( f|_W)$.  For simple functions $G$, like polynomials, the derivative $G^{\prime}$ can be easily calculated with the limit definition for the real numbers $\RsubD(W)$, the result will be the same as we get using infinitesimals. The derivative of the function $G$ at $ f|_{W} $ is $G'( f|_W)$. 

If $G$ is a continuous function: $\RsubD(X) \to \RsubD(X)$, that can send a number defined to extent $W$ to a number defined to extent $U \ne W$, then the derivative of $G$ at the point given by the continuous real-valued function $f|_{W}$ for $W \in \mathcal O(X)$ is given by 
\begin{equation}
G( f|_{W} + d f|_{W}) = G(f|_{W}) +  f|_{\partial W}\cdot G'( f|_{ W})
\end{equation}
where $d f|_{W} =  f|_{\partial W}$ with $ \partial W$ the boundary of $W$.

A Dedekind manifold $M_D(X)$ is modelled on the sheaf of Dedekind real numbers  $\RsubD(X)$ in the sense that there is an atlas of charts for $M_D(X)$. A chart for $M_D(X)$ consists of a subsheaf 
$S(U)$ of $M_D(X)$ together with a 1:1 function $\phi$ that maps $M_D(U)$ onto an open subset of  $\RsubD(X)$. An atlas for the sheaf $M_{D}(X)$ is a collection of mutually compatible charts $\{M_D((U_i) , \phi_i);  i \in I \}$ such that $\cup_{i\in I}M_D( U_i ) = M_{D}(X)$. The charts $(M_D(U) , \phi)$ and $(M_D(V), \psi)$ are compatible  if when $U \cap V \ne \emptyset$ the sets $\phi(M_D(U \cap V))$ and  $\psi(M_D(U \cap V))$ are open subsets of $\RsubD(X)$ and the overlap map $ \phi \circ \psi^{-1}$ is a diffeomorphism between these two subsheafs. The charts are $C^r$ compatible when the overlap map is a diffeomorphism of class $C^r$ for $1 \leq r \leq \infty$.



\begin{thebibliography}{99}






\bibitem{isham} Isham C.J. and Butterfield, J. "Some possible Roles for
Topos Theory in Quantum Theory and Quantum Gravity", Fndns of Physics
(2000)
\bibitem{bell} J.S.Bell, {\em Speakable and Unspeakable in Quantum
Mechanics"}, pp117-118 (Cambridge University Press,1993)\smallskip

\bibitem{jlbell} J.L.Bell, {\em A Primer of Infinitesimal Analysis"}, pp117-118 (Cambridge University Press,1998)\smallskip

\bibitem{bitbol} M. Bitbol, ``Are there particles and quantum jumps?'',
in R. Nair (ed),{\em Mind, Matter and Mystery}, pp52-53,Scientia, New
Delhi, (2001). \smallskip
 
\bibitem{bohr} N. Bohr, ``Ueber die Antwendung der Quantentheorie auf
den Atombau. I. Die Grundpostulate der Quantentheorie'', Z. Phys., 13
(1923) p118, {\footnotesize\noindent ``one is obliged \dots to always
keep in mind the domain of application'' } \linebreak
and {\em Atomic Theory and the Description of Nature}, Cambridge (1932),
p73,  {\footnotesize\noindent `` constantly keep the possibilities of
definition as well as of observation before the mind''} \smallskip 

\bibitem{cohen-tannoudji}C. Cohen-Tannoudji, B. Diu and F. Laloe, {\em
Mecanique quantique},(Hermann, Paris, 1977).\smallskip 

\bibitem{conway}J.H.Conway,{\em On Numbers and Games}, Academic Press, (1976)

\bibitem{dacosta} N.C.A. da Costa, D. Krause, S. French,``The
Schroedinger Problem'',in M. Bitbol, O. Darrigol (ed) {\em Erwin
Schroedinger, Philosophy and the Birth of Quantum Mechanics}, pp450-453,
Editions Frontieres, Gif-sur-Yvette, France. \smallskip
\bibitem{cush} Cushing J.T. (1987):``Foundational Problems in and Methodological Lessons from Quantum Field Theory'', in H. R. Brown and R. Harr\'e (eds.), Philosophical Foundations of Quantum Field Theory, Claredon Press, Oxford (1988)\smallskip 
\bibitem{dedekind} R. Dedekind, "Was sind und was sollen die
Zahlen",Braunschweig, 1887, SIII.

\bibitem{dirac} P.A.M. Dirac, " The Principles of Quantum Mechanics",4th
ed.,Oxford University Press (1958) p 36.

\bibitem{disalle} R. DiSalle, " Spacetime theory as physical geometry " {\em Erkenntnis},
\underbar{42}, 317-337 (1995)

\bibitem{einstein1} Einstein, A. "Johannes Kepler", Frankfurter Zeitung ,9
November,1930, on the 300th anniversary of Kepler's death, republished
in "Ideas and Opinions", Crown Publishers, Inc., (1954), pp 256-259.

\bibitem{bachelard} G. Bachelard, {\em Le nouvel esprit scientifique},
pp 129-138 (Presses Universitaires de France, 1934; "Quadrige":
1995).

\bibitem{einstein2} A. Einstein, "Geometry and Experience", Lecture before
the Prussian Academy of Sciences, 27 January, 1921,  republished
in "Ideas and Opinions", Crown Publishers, Inc., (1954) pp 227-232.

\bibitem{glaub} R.J.Glauber,Phys.Rev.\underbar{131},2766 (1963), S.T. Ali,
 J-P. Antoine, J-P. Gazeau, {\em Coherent States, Wavelets and Their Generalizations}, Springer-Verlag New York, Inc., (2000).

\bibitem{haag} R. Haag and D. Kastler, ``An Algebraic Approach to Quantum Field Theory '', J. Math. Phys. \underbar{5}, 215 (1964). \smallskip

\bibitem{hepp} K. Hepp, ``The Classical Limit for Quantum Mechanical
Correlation Functions '', Commun. Math. Phys. \underbar{35}, 265 -
277(1974). 
\bibitem{heunen} Heunen, C. and Spitters, B. (2007) `A topos for algebraic quantum
theory', {\it arXiv:0709.4364.}
\bibitem{home} D. Home, {\em Conceptual Foundations of Quantum Physics}, Plenum Press, 
New York and London, (1997) \smallskip
\bibitem{isham1} Isham, C.J. and Butterfield, J. (2000) `Some possible Roles for
Topos Theory in Quantum Theory and Quantum Gravity'.{\it  Foundations of Physics, 30}, pp. 1707-1735.\smallskip
\bibitem{isham2} Doering, A. and  Isham, C.J. (2007) `A topos foundation for theories of physics: I. Formal languages for physics',{\it  arXiv:quant-ph/0703060};  `II. Daseinisation and the liberation of quantum theory'. {\it arXiv:quant-ph/0703062} ;  `III. The representation of physical quantities with arrows'.{\it arXiv:quant-ph/0703064};  `IV. Categories of systems'. {\it arXiv:quant-ph/0703066}.

\bibitem{jauch} J. M. Jauch, {\em Foundations of Quantum Mechanics},
p275-280, Addison-Wesley, Reading Massachusetts, (1968) \smallskip

\bibitem{lawvere}W. Lawvere,  ``Quantifiers as Sheaves'', Actes Congres
Intern. Math. {\bf 1}, (1970).
\bibitem{kato}T. Kato, \emph{ Perturbation Theory for Linear Operators}
(Springer-Verlag, New York, 1966).

\bibitem{kirk}Kirkup L. and Frenkel R.B. {\em An Introduction to Uncertainity in Measurement using the GUM ( Guide to the Expression of Uncertainity in Measurement},Cambridge University Press(2006)
\smallskip  
\bibitem{kock}A. Kock,  {\em Synthetic Differential Geometry}, Cambridge University Press,
Cambridge (1981).

\bibitem{mack}G.W.Mackey,  {\em Induced Representations of Groups and Quantum Mechanics}, W.A.Benjamin, INC., New York (1968).
\bibitem{maclane1} S. MacLane {\em Categories for the Working Mathematician} 2nd edition(Springer--Verlag, New York, 1998).
\bibitem{maclane}S. MacLane and I. Moerdijk, {\em Sheaves in Geometry
and Logic} (Springer--Verlag, New York, 1994).
\bibitem{johnstone} P. T. Johnstone {\em Topos Theory} (Academic Press, New York, 1977).
\bibitem{reyes}I. Moerdijk and G.E. Reyes, {\em Models for Smooth Infinitesimal Analysis} (Springer--Verlag, New York, 1991).

\bibitem{powers}R. T. Powers,{\em Self-adjoint algebras of unbounded
operators}, Commun. Math. Phys. \underbar{21},pp 85-124, (1971), 

\bibitem{stout}L. N. Stout, Cahiers Top. et Geom. Diff. {\bf XVII},
295 (1976).

\bibitem{adelman1} M. Adelman and J. V. Corbett, Applied Categorical
Structures {\bf 3}, 79 (1995).
\bibitem{adelman2} M.Adelman and J.V. Corbett: "Quantum Mechanics
as an Intuitionistic form of Classical Mechanics"
Proceedings of the Centre Mathematics and its Applications, pp15-29, ANU,
Canberra (2001).\smallskip

\bibitem{arxives}J. V. Corbett and T. Durt, ``Quantum Mechanics interpreted in Quantum Real Numbers'', 
quant-ph/0211180, 1-26, (2002).
\bibitem{durt2} J.V.Corbett and T Durt, {\em Collimation processes in quantum mechanics 
interpreted in quantum real numbers}, Studies in History and Philosophy of Modern Physics, {\bf 40}, pp 68-83, (2009).
\bibitem{arxives2}J. V. Corbett and T. Durt, ``Quantum Mechanics as a Space-Time Theory'', 
quant-ph/051220, 1-27, (2005).

\bibitem{swan}R.G. Swan, {\em The Theory of Sheaves} (University of
Chicago Press, Chicago and London, 1964)
\S3.2, pp29-31.
\bibitem{vonneumann} J. V. von Neumann: {\em Mathematische grundlagen
der Quanten-Mechanik },  Springer-Verlag, Berlin (1932). English
translation: {\em Mathematical Foundations of Quantum Mechanics},
Princeton University Press, Princeton, 1955.\smallskip
\bibitem{weyl} H. Weyl: {\em Philosophy of Mathematics and Natural
Science}, Princeton University Press, Princeton
(1949).\smallskip 
\bibitem{inoue} A. Inoue,{\em Tomita-Takesaki Theory in Algebras of 
Unbounded Operators} Lecture Notes in Mathematics; 1699, Springer (1999),
K. Schmudgen, {\em Unbounded Operator Algebras and Representation Theory}
Operator Theory: Vol. 37, Birkhauser Verlag (1990).\smallskip
\bibitem{kothe} G. K\"othe,{\em Topological Vector spaces } Vol II, Section 41,4,  Springer- Verlag, Berlin,(1979).\smallskip
\bibitem{jaek} M-T. Jaekel and S.Reynaud,{\em Conformal symmetry and quantum relativity}
Found. of Phys. \em{28},439 (1998)\smallskip
\bibitem{james} G.J.O.Jameson, {\em Topology and Normed Spaces}, Chapman and Hall, London, (1974). \smallskip
\bibitem{jam} M. Jammer, {\em The Conceptual Development of Quantum
Mechanics}, p175 NewYork (1966) \smallskip
\bibitem{johnstone} P.T.Johnstone,{\em Topos Theory}, pp210-220, Academic Press, New York, (1977)
\bibitem{levy} J.-L. L\'{e}vy-Leblond, {\em Galilei group and Galilean
invariance}, in E.M. Loebl (ed), {\em Group Theory and Its Applications},
Academic Press, New York, pp221-299, (1971). \smallskip
\bibitem{sanch} D.A.Sanchez,  {\em Ordinary differential equations and
stability theory : An introduction},  pp139-141, W.H. Freeman and Company,
San Francisco, (1968) or E.L.Ince, {\em Ordinary differential equations}, Chapter 3, Dover Publications, USA, (1956).  
\bibitem{saun} S. Saunders,{\em The Algebraic Approach to Quantum Field Theory }, in {\em Philosophical Foundations of Quantum Field Theory}, edited by H.R. Brown and R. Harr\'{e}, Oxford University Press,
Oxford, (1988)
\bibitem{reed} M. Reed and B. Simon, {\em Methods of Modern Mathematical 
Physics} Vol. 1, Functional Analysis, Academic Press, New York and London,
(1972). \smallskip
\bibitem{wigner} E.P.Wigner, {\em On Unitary Representations of the Inhomogeneous Lorentz Group}. Ann. of Math. \em{40}(1939) 149.



\end{thebibliography}

\end{document}